\documentclass[prc,twocolumn,aps,tightenlines,showpacs,floatfix]{revtex4}
\usepackage{bm}
\usepackage{color}
\usepackage{graphicx,epsfig}
\usepackage{longtable}
\usepackage{xcolor}
\usepackage{color}


\usepackage{dcolumn}     %
\newcommand{\nuc}[2] {$^{#1}$#2}

\def\beq{\begin{equation}}
\def\eeq{\end{equation}}
\def\beqy{\begin{eqnarray}}
\def\eeqy{\end{eqnarray}}

\begin{document}

{
\title{Chiral Effective Field Theory Calculations of Weak  Transitions in Light Nuclei}
\author{G.~B. \ King$^{\rm a}$, L.\ Andreoli$^{\rm a}$, S.\ Pastore$^{\rm a,b}$, M.\ Piarulli$^{\rm a}$, 
 R.\ Schiavilla$^{\rm c,d}$, R.~B.\ Wiringa$^{\rm e}$, J.\ Carlson$^{\rm f}$, and S.\ Gandolfi$^{\rm f}$}
\affiliation{
$^{\rm a}$\mbox{Department of Physics, Washington University in Saint Louis, Saint Louis, MO 63130, USA}
$^{\rm b}$\mbox{McDonnell Center for the Space Sciences at Washington University in St. Louis, MO 63130, USA}
$^{\rm c}$\mbox{Department of Physics, Old Dominion University, Norfolk, VA 23529, USA}
$^{\rm d}$\mbox{Theory Center, Jefferson Lab, Newport News, VA 23606, USA}
$^{\rm e}$\mbox{Physics Division, Argonne National Laboratory, Argonne, Illinois, IL 60439, USA}
$^{\rm f}$\mbox{Theoretical Division, Los Alamos National Laboratory, Los Alamos, NM 87545, USA }
}
\date{\today}
\begin{abstract}
We report Quantum Monte Carlo calculations of weak transitions in $A\leq 10$ nuclei,
based on the Norfolk two- and three-nucleon chiral interactions, and associated
one- and two-body axial currents. We find that the contribution from two-body currents is 
at the 2--3\% level, with the exception of matrix elements entering the 
rates of $^8$Li, $^8$B, and $^8$He beta decays. These
matrix elements are suppressed in impulse approximation based on the (leading
order) Gamow Teller transition operator alone; two-body currents
provide a $20$--$30\%$ correction, which is,
however, insufficient to  bring theory in agreement with experimental data. 
For the other transitions, the agreement with the data is satisfactory, and the 
results exhibit a negligible to mild model dependence when different combinations
of Norfolk interactions
are utilized to construct the nuclear wave functions.  We report a complete
study of two-body weak transition densities which reveals the expected universal behavior
of two-body currents at short distances throughout the range of $A\,$=$\,3$ to $A\,$=$\,10$ systems
considered here.  
\end{abstract}

\maketitle
}
\section {Introduction}
\label{sec:intro}

In this work, we present Quantum Monte Carlo (QMC) calculations, including both variational 
Monte Carlo (VMC) and Green's function Monte Carlo (GFMC) calculations,
of Gamow-Teller (GT) matrix elements
entering beta-decay and electron-capture rates in $A\,$=$\,3$--$10$ nuclei. 
These observables are experimentally known (in most 
cases) at the sub-percent level, and are used here primarily 
to validate our microscopic theoretical modeling of the 
nucleus as a system of nucleons interacting with each other via
effective interactions, and with electroweak
probes via effective currents.
Specifically, this modeling is based on local two- and
three-nucleon interactions formulated in configuration space, and derived
from a chiral effective field theory ($\chi$EFT) that retains,
in addition to nucleons and pions, $\Delta$-isobars as explicit degrees of
freedom~\cite{Piarulli:2014bda,Piarulli:2016vel,Baroni:2016xll,Baroni:2018fdn}.
They are referred to below as the Norfolk interactions and are denoted as NV2+3.
Accompanying these interactions are one- and two-body axial currents---local in configuration
space---derived within the same $\chi$EFT
formulation~\cite{Baroni:2015uza,Baroni:2016xll,Baroni:2018fdn}.  In the calculations to follow,
we include up to tree-level contributions at next-to-next-to-next-to-leading order (N3LO) in the 
chiral expansion, and disregard subleading corrections involving loops and higher order contact terms. 

An analogous study was recently reported by some
of the present authors in Ref.~\cite{Pastore:2017uwc}.
There, the QMC calculations were based on the Argonne-$v_{18}$ (AV18)
two-nucleon~\cite{Wiringa:1994wb} and Illinois-7 (IL-7)
three-nucleon~\cite{doi:10.1063/1.2932280} interactions, 
in combination with the axial currents of Ref.~\cite{Baroni:2015uza}.
We found agreement with the experimental Gamow-Teller matrix elements
at the $\approx 2\%$ level 
for the $A\,$=$\,6$ and $7$ systems, and at the $\approx 10\%$ level
in $^{10}$C. The large uncertainty in the $A\,$=$\,10$ 
transition is primarily systematic and results from the narrow energy 
separation between the first two $J^\pi\,$=$\,1^+$ states in $^{10}$B, which 
makes it hard to precisely disentangle them~\cite{Pastore:2017uwc}.  
The study of Ref.~\cite{Pastore:2017uwc} found that two-body
currents generate an additive contribution of the order of $\approx 3\%$ 
and concluded that the agreement with the data is mainly attributable to 
the use of fully correlated nuclear wave functions, rather than two-body 
effects in the currents. 

In the meantime, no-core shell-model calculations of weak matrix 
elements based on chiral interactions and currents~\cite{Gysbers:2019uyb}
found the sign of the overall correction generated by two-body 
currents to be opposite to that obtained in Ref.~\cite{Pastore:2017uwc} for the same systems 
(but in agreement with a hybrid calculation of the $A\,$=$\,6$ decay reported in Ref.~\cite{Gazit:2009kz}).
This discrepancy was attributed to the hybrid nature 
of the calculation of Ref.~\cite{Pastore:2017uwc}, {\it i.e.},
to the mismatch between the two- and three-body
correlations implemented to construct the nuclear wave 
functions---and induced by the AV18 and IL7---and 
those entering the axial currents which were instead 
derived from $\chi$EFT.

In this work, by reexamining the evaluation of these weak matrix
elements with the NV2+3 chiral interactions~\cite{Piarulli:2014bda,Piarulli:2016vel,Baroni:2016xll,Baroni:2018fdn},
in combination with consistent chiral axial currents at tree-level~\cite{Baroni:2015uza},
we aim to address and explore the aforementioned claim.
We investigate the sensitivity of the calculated matrix elements with 
respect to different choices of regulators and to different 
strategies adopted to constrain the three-body Norfolk interactions (NV3). 
This latter aspect is important in order to understand the interplay between these
interactions and the axial currents, since the strength of the contact current and that of
the three-body interaction of one-pion-range are rigorously related to each other by the symmetries
imposed in the $\chi$EFT formulation~\cite{Gardestig:2006hj,Gazit:2008ma,Schiavilla:2017}.

This study has several merits. First, we report new GFMC results of the 
energy spectra of $A\le10$ nuclei based on 
two classes of NV2+3 interactions. In addition to the systems
studied in Ref.~\cite{Pastore:2017uwc}, we study weak transitions in
$A\,$=$\,8$ nuclei where we find that two-body axial currents provide 
a large correction to the one-body results. Finally, we provide the first 
calculations of two-body weak transition densities which shed light
on the role of short-range physics in these observables. 

Searches for physics beyond the Standard Model (BSM) via beta 
decay are the focus of current and planned experimental 
programs carried out at the Facility for Rare Isotope Beams (FRIB), the 
the University of Washington, and Argonne National Laboratory (see, {\it e.g.},
Ref.~\cite{Gonzalez-Alonso:2018omy} and references therein). 
Among the targets under consideration are  $^6$He, $^8$Li, $^8$B
and $^{10}$C.  A systematic study of axial-current matrix elements in these systems
is a prerequisite for all further investigations and BSM
searches in beta decay.  Beta decays are ideal processes to 
assess the validity of the dynamical inputs of {\it ab inito} calculations,
namely many-body correlations and weak currents. The latter
also impact calculations of neutrinoless double beta decay matrix
elements, whose knowledge is critical to the neutrinoless double beta 
decay experimental program~\cite{Engel:2016xgb}. 

This paper is structured as follows: a brief review of the 
QMC computational method and Norfolk interactions is given in 
Secs.~\ref{sec:qmc} and~\ref{sec:hamiltonian}. The many-body axial 
currents used in this work are reported in Sec.~\ref{sec:axial}.
Results and conclusions are provided in Secs.~\ref{sec:res}
and~\ref{sec:concl}.

\section {Quantum Monte Carlo method}
\label{sec:qmc}

The Quantum Monte Carlo methods used in this study have been
described in detail in several review articles, the most recent of which being
Refs.~\cite{Carlson:2014vla,Lynn:2019rdt}. Here, we sketch the computational 
procedure and refer the interested reader to Refs.~\cite{Carlson:2014vla,Lynn:2019rdt} and references therein.

We seek accurate solutions of the many-nucleon Schr\"{o}dinger equation
\beqy H \Psi(J^\pi;T,T_z)= E \Psi(J^\pi;T,T_z) \ ,\eeqy
where $J^\pi$ are the total angular momentum and parity of the state, and
$T$ and $T_z$ are the total isospin and its projection, respectively.  
We use the Hamiltonian
\beqy \label{eq:hamiltonian} H = \sum_{i} K_i + {\sum_{i<j}} v_{ij} + \sum_{i<j<k}
V_{ijk} \ ,
\eeqy
where $K_i$ is the non-relativistic kinetic energy, and $v_{ij}$ and $V_{ijk}$
are the NV2 and NV3 local chiral 
interactions~\cite{Piarulli:2014bda, Piarulli:2016vel,Baroni:2016xll,Baroni:2018fdn},
collectively denoted as NV2+3.

The VMC trial function $\Psi_V(J^\pi;T,T_z)$ for a given nucleus is constructed
from products of two- and three-body correlation operators acting on an
antisymmetric single-particle state of the appropriate quantum numbers.
The correlation operators are designed to reflect the influence of the
interactions at short distances, while appropriate boundary conditions
are imposed at long range~\cite{Wiringa:1991kp,Pudliner:1997ck,Nollett:2000ch,Nollett:2001ub}.
The $\Psi_V(J^\pi;T,T_z)$ contains variational parameters
that are adjusted to minimize the expectation value
\begin{equation}
 E_V = \frac{\langle \Psi_V | H | \Psi_V \rangle}
            {\langle \Psi_V   |   \Psi_V \rangle} \geq E_0 \ ,
\label{eq:expect}
\end{equation}
which is evaluated by Metropolis Monte Carlo integration~\cite{Metropolis:1953am}.
The lowest value for $E_V$ is then taken as the approximate ground-state energy of
the exact lowest eigenvalue of $H$, $E_0$, for the specified quantum numbers.

A good trial wave function is given by
\begin{equation}
   |\Psi_V\rangle =
      {\cal S} \prod_{i<j}^A
      \left[1 + U_{ij} + \sum_{k\neq i,j}^{A}\widetilde{U}^{TNI}_{ijk} \right]
      |\Psi_J\rangle.
\label{eq:psit}
\end{equation}
The Jastrow wave function $\Psi_J$ is fully antisymmetric and has the
$(J^\pi;T,T_z)$ quantum numbers of the state of interest, while $U_{ij}$
and $\widetilde{U}^{TNI}_{ijk}$ are two- and three-body correlation
operators. 

The GFMC method~\cite{Carlson:2014vla} improves on the VMC wave functions by acting
on $\Psi_V$ with the operator $\exp \left[-\left(H - E_0\right)\tau\right]$.
The operator is applied in a sequence of small imaginary-time steps $\Delta\tau$ to produce a 
propagated wave function
\begin{eqnarray}
 \Psi(\tau) = e^{-({H}-E_0)\tau} \Psi_V
          = \left[e^{-({H}-E_0)\triangle\tau}\right]^{n} \Psi_V \ .
\end{eqnarray}
Obviously $\Psi(\tau\!=\!0) \,$=$\, \Psi_V$ and $\Psi(\tau \rightarrow \infty)\,$=$\, \Psi_0$.
Quantities of interest are evaluated in terms of a ``mixed'' expectation value between 
$\Psi_V$ and $\Psi(\tau)$: 
\begin{eqnarray}
\langle O(\tau) \rangle_M & = & \frac{\langle \Psi(\tau) | O |\Psi_V
\rangle}{\langle \Psi(\tau) | \Psi_V\rangle},
\label{eq:expectation}
\end{eqnarray}
where the operator $O$ acts on the trial function $\Psi_V$.
The desired expectation values would, of course, have $\Psi(\tau)$ on both
sides; by writing $\Psi(\tau) = \Psi_V + \delta\Psi(\tau)$  and neglecting
terms of order $[\delta\Psi(\tau)]^2$, we obtain the approximate expression
\begin{eqnarray}
\langle O (\tau)\rangle &=&
\frac{\langle\Psi(\tau)| O |\Psi(\tau)\rangle}
{\langle\Psi(\tau)|\Psi(\tau)\rangle}  \nonumber \\
&\approx& \langle O (\tau)\rangle_M
    + [\langle O (\tau)\rangle_M - \langle O \rangle_V] ~,
\label{eq:pc_gfmc}
\end{eqnarray}
where $\langle O \rangle_{\rm V}$ is the variational expectation value.

For off-diagonal matrix elements required by the transitions
we are interested here, the generalized mixed estimate is given by the expression
\begin{eqnarray}
&& \frac{\langle\Psi^f(\tau)| O |\Psi^i(\tau)\rangle}{\sqrt{\langle \Psi^f(\tau) | \Psi^f(\tau)\rangle}
\sqrt{\langle \Psi^i(\tau) |\Psi^i(\tau)\rangle}} \nonumber \\
&\approx&
  \langle O(\tau) \rangle_{M_i}
+ \langle O(\tau) \rangle_{M_f}-\langle O \rangle_V \ ,
\label{eq:extrap}
\end{eqnarray}
where
\begin{eqnarray}
\langle O(\tau) \rangle_{M_f} 
& = & \frac{\langle \Psi^f(\tau) | O |\Psi^i_V\rangle}
           {\langle \Psi^f(\tau)|\Psi^f_V\rangle}
      \sqrt{\frac{\langle \Psi^f_V|\Psi^f_V\rangle}
           {\langle \Psi^i_V | \Psi^i_{V}\rangle}} \ , 
\label{eq:mixed_f} 
\end{eqnarray}
and $\langle O(\tau) \rangle_{M_i}$ is defined similarly.
For more details see Eqs.~(19)--(24) and the accompanying discussions in Ref.~\cite{Pervin:2007sc}.

\section{Norfolk interaction models}
\label{sec:hamiltonian}
We base our calculations of weak transitions in $A$=6--10 
on the local NV2 and NV3 interactions developed in 
Refs.~\cite{Piarulli:2014bda,Piarulli:2016vel,Baroni:2016xll,Baroni:2018fdn}.
The NV2 model has been derived from a $\chi$EFT that uses pions, 
nucleons and $\Delta$'s as fundamental degrees of freedom. It consists of 
a long-range part, $v_{ij}^{\rm L}$, mediated by one- and two-pion exchanges, 
and a short-range part, $v_{ij}^{\rm S}$, described in terms of contact 
interactions with strengths specified by unknown low-energy constants
(LECs).  The strength of the long-range part is fully determined by the
nucleon and nucleon-to-$\Delta$ axial coupling constants $g_A$ and $h_A$, 
the pion decay amplitude $F_\pi$, and the LECs $c_1$, $c_2$, $c_3$, $c_4$, 
and $b_3+b_8$, constrained by reproducing $\pi N$ scattering data~\cite{Krebs:2007rh}. 
The LECs entering the contact interactions are fixed by fitting
nucleon-nucleon scattering data from the most recent and up-to-date database 
collected by the Granada group~\cite{Perez:2013jpa,Perez:2013oba,Perez:2014yla}.
The value for $h_A$ is taken from the large $N_c$ expansion or strong-coupling 
model~\cite{Green:1976wx}. The value of the nucleon axial coupling constant
used to construct the nucleon-nucleon interaction accounts for the Goldberger-Treiman 
discrepancy~\cite{Arndt:1994bu,Stoks:1992ja} and, to distinguish it from the 
experimental value of $g_A\,$=$\,1.2723\, (23)$~\cite{PDG} entering the axial currents, we denote it 
with $\tilde{g}_A$.  For completeness, Tables~\ref{tab:lec1} and~\ref{tab:lec2}
report the values of these constants, along with the pion and nucleon masses,
the $\Delta$-nucleon mass difference,
the electron mass, and the fine structure constant $\alpha$
used in the NV2 interactions (these last two characterize the electromagnetic part of the NV2s~\cite{Piarulli:2014bda}).
	\begin{table}
		\begin{tabular}{cccccccc}
			\hline\hline
			$\tilde{g}_A$  & $h_A$ & $F_{\pi}$ & $c_1$ & $c_2$ & $c_3$ & $c_4$ &  $b_3+b_8$ \\
			\hline
			$1.29$ &$2.74$ & $184.80$ &$-0.57$ &$-0.25$ &$-0.79$ & $1.33$&  $1.40$ \\
			\hline\hline
		\end{tabular}
		\caption{Values of (fixed) low energy constants (LECs) used in this work: $\tilde{g}_A$ and $h_A$ are
			adimensional, $F_\pi\,$=$\,2f_{\pi}$ is given in MeV, and the remaining LECs are given in GeV$^{-1}$.
			See text for explanation.}
		\label{tab:lec1}
	\end{table}
\begin{table*}
	\begin{tabular}{cccccccc}
		\hline\hline
		$m_{\pi_0}$ &$m_{\pi_{\pm}}$ & $M_n$ &$M_p$ &$m_{\Delta N}$& $m_e$ & $\alpha^{-1}$ \\
		\hline
		$134.9766$& $139.5702$ & $939.56524$  & $938.27192$&$293.1$ & $0.510999$ &  $137.03599$ \\
		\hline\hline
	\end{tabular}
	\caption{Values of charged and neutral pion masses, proton and neutron masses, $\Delta$-nucleon
		mass difference, and electron mass (all in MeV), and of the (adimensional) fine structure
		constant $\alpha$.  Note that $\hbar c$ is taken as 197.32697 MeV$\,$fm.}
	\label{tab:lec2}
\end{table*}

The contact terms are implemented using a Gaussian representation
of the three dimensional delta function, with $R_S$ denoting the Gaussian 
parameter~\cite{Piarulli:2016vel,Piarulli:2014bda,Baroni:2016xll,Baroni:2018fdn}. 
The pion-range operators are strongly singular at short range in configuration space,
and are regularized by a radial function characterized
by a cutoff $R_L$~\cite{Piarulli:2016vel,Piarulli:2014bda,Baroni:2016xll,Baroni:2018fdn}.
There are two classes (I and II) of NV2s, differing only in the range of energy over
which they are fitted to the database---class I up to 125 MeV, and
class II up to 200 MeV. 
For each class, two combinations of short- and long-range regulators 
have been used, namely ($R_S$, $R_L$)=(0.8, 1.2) fm (models NV2-Ia and NV2-IIa) and 
($R_S$, $R_L$)=(0.7, 1.0) fm (models NV2-Ib and NV2-IIb).
Class I (II) fits about 2700 (3700)
data points with a $\chi^2$/datum $\lesssim 1.1$ 
($\lesssim 1.4$)~\cite{Piarulli:2016vel,Piarulli:2014bda}. 

 \begin{figure}
 \vspace*{.2in}
 \includegraphics[width=2.2in]{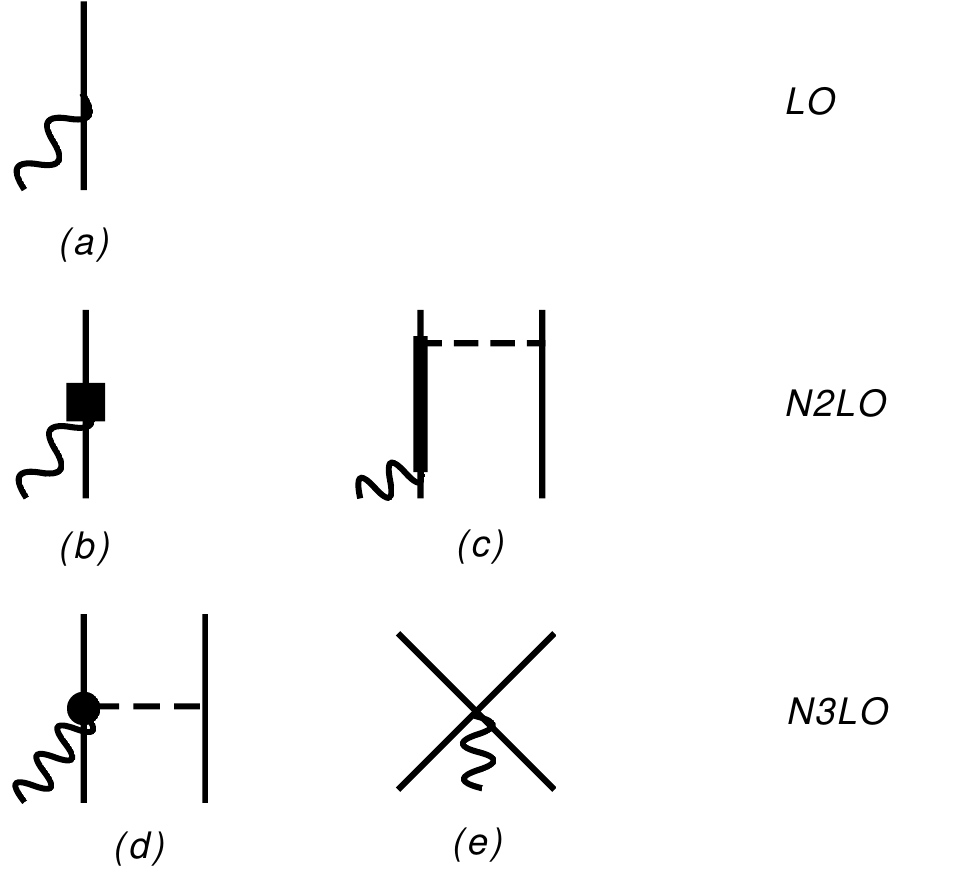}
 \caption{Diagrams illustrating the contributions to the axial current up
to N3LO used in this work.  Nucleons, $\Delta$-isobars, pions, 
and external fields are denoted by solid, thick-solid, dashed, and wavy lines,
respectively.  The square in panel (b) represents relativistic
corrections, while the dot in panels (d) denotes a vertex induced by subleading
terms in the $\pi$-nucleon chiral Lagrangian~\cite{Baroni:2016xll}.}
\label{fig:currax}
\end{figure}

The NV2 models were found to provide insufficient attraction in GFMC calculations
of the binding energies of light nuclei~\cite{Piarulli:2016vel}. To
remedy this shortcoming, a consistent three-body interaction was constructed
up to N2LO in the chiral 
expansion. It consists of a long-range part mediated by two-pion 
exchange and a short-range part parametrized in terms of two contact interactions~\cite{vanKolck:1994yi,Epelbaum:2002vt} proportional to the LECs $c_D$ and $c_E$.
These LECs have been obtained by fitting either
observables that involve exclusively strong interactions~\cite{Lynn:2015jua,Tews:2015ufa,Lynn:2017fxg,Piarulli:2017dwd}
or a combination of observables that involve both strong and weak interactions~\cite{Gazit:2008ma,Marcucci:2011jm,Baroni:2018fdn}. 
This last strategy is feasible because of the relation established
in $\chi$EFT~\cite{Gardestig:2006hj} that links $c_D$ with the LECs entering the contact axial current at N3LO~\cite{Gazit:2008ma,Marcucci:2011jm,Schiavilla:2017}
(see next section for details).

In Ref.~\cite{Piarulli:2017dwd}, $c_D$ and $c_E$ were determined by simultaneously reproducing the experimental trinucleon ground-state energies and $nd$ doublet 
scattering length. These first-generation
NV2+3 interactions, denoted with NV2+3-Ia/b and NV2+3-IIa/b,
have been implemented in both VMC and GFMC codes and
used to study static properties of light 
nuclei~\cite{Piarulli:2016vel,Piarulli:2017dwd,Lynn:2019rdt,Cirigliano:2019vdj,10.3389/fphy.2019.00245,Gandolfi:2020pbj}, and 
in auxiliary-field diffusion Monte Carlo (AFDMC)~\cite{Schmidt:1999lik}, Brueckner-Bethe-Goldstone (BBG)~\cite{bbg1,bbg2} and 
Fermi hypernetted chain/single-operator chain (FHNC/SOC)~\cite{FR75,PW79}
approaches to investigate the equation of state of neutron matter~\cite{Piarulli:2019pfq,Bombaci:2018ksa}.

In more recent work~\cite{Baroni:2018fdn}, $c_D$ and $c_E$ were constrained by fitting, in addition to the trinucleon energies,
the empirical value of the GT matrix element in tritium $\beta$ decay. 
These second-generation NV2+3 interactions were designated as NV2+3-Ia$^*$/b$^*$ and NV2+3-IIa$^*$/b$^*$. 
These two different procedures for fixing $c_D$ and $c_E$ produced
rather different values for these LECs.  They are reported in Table~\ref{tab:cdce}.

\begin{center}
	\begin{table}
	\resizebox{245pt}{!}{
		\begin{tabular}{l|r r r r}
			\hline\hline
			       & Ia  (Ia*)         &        Ib (Ib*)   &      IIa (IIa*) & IIb (IIb*)\\
			\hline
			$c_D$  &    3.666 (--0.635)  &  --2.061 (--4.71) &   1.278 (--0.61)    & --4.480 (--5.25) \\
			$c_E$  &  --1.638 (--0.090)  &  --0.982 (  0.55) & --1.029 (--0.35)    & --0.412 (0.05)\\
			$z_0$  &    0.090 (  1.035)  &    2.013 (2.881)  &   0.615 (1.03)      &  2.806 (3.059)   \\
				\hline\hline
		\end{tabular}
		}
		\caption{Adimensional $c_D$ and $c_E$ values of the contact terms in the NV3
			interactions obtained from fits to  {\it i}) the $nd$ scattering length and 
			trinucleon binding energies~\cite{Piarulli:2017dwd}; and {\it ii}) the 
			central value of the $^3$H GT matrix element and the trinucleon binding energies (starred
			values)~\cite{,Baroni:2018fdn}.  The adimensional $z_0$ values are obtained using the relation given
			in Eq.~(\ref{eq:ez0}).
				}
		\label{tab:cdce}
	\end{table}
\end{center}
\section{$\chi$EFT axial currents}
\label{sec:axial}

\begin{figure}%
\includegraphics[height=3.3in]{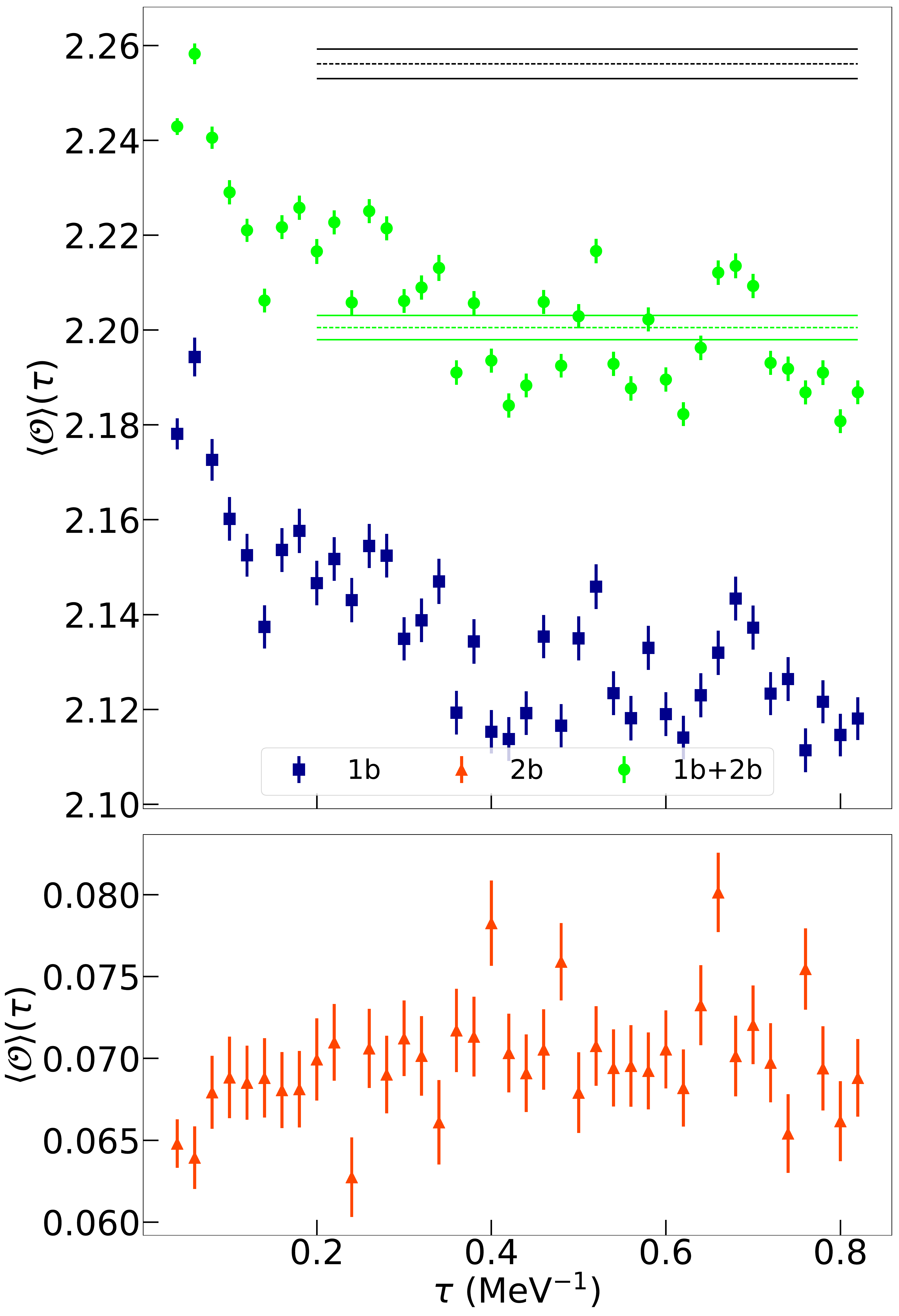}
\caption{(Color online)   Propagation of the $^6$He$\,\rightarrow\,^6$Li transition matrix element 
as function of imaginary time $\tau$, based on the NV2+3-Ia Hamiltonian.
Results with the one-body current at LO and currents beyond LO are indicated with 1b
and 2b, respectively.  Dashed and solid lines represent central values and associated error bars. Black dashed
and solid lines denote the VMC results. See text for further explanations.}
\label{fig:it}
\end{figure}

Many-body axial currents have been first examined within $\chi$EFT 
by Park and collaborators in Ref.~\cite{Park:1993jf}. In that work,
the authors retained pions and nucleons in their effective theory
and calculated the two-body axial currents up to one-loop terms.
The derivation neglected, for example,
pion-pole contributions. More recently, two-body axial currents
with pions and nucleons have been derived by the Bonn group~\cite{Krebs:2016rqz} 
using the unitary transformation method, and by the JLab-Pisa group using
time-ordered perturbation theory~\cite{Baroni:2015uza,Baroni:2016xll}. 
The two derivations differ in the treatment of reducible diagrams. 
When calculating box diagrams entering the electromagnetic charge and current
operators~\cite{Pastore:2008ui,Pastore:2009is,Pastore:2011ip,Kolling:2009iq,Kolling:2011mt}, 
the two methods lead to results that are in agreement. However, as discussed
at length in Refs.~\cite{Baroni:2016xll,Krebs:2020rms}, the two groups find
different results for the box diagrams in the two-body axial current 
operator at N4LO. The numerical impact of this difference has been
investigated in Refs.~\cite{Baroni:2016xll,Baroni:2018fdn}, where both the JLab-Pisa
and Bonn versions of the N4LO current operators have been implemented to calculate the GT
matrix element in triton beta decay. In those studies, it was found that
the corrections generated by the JLab-Pisa and Bonn N4LO operators 
are qualitatively in agreement (they both quench the GT matrix element
at leading order), and provide, respectively, a $\approx 6\%$ and $\approx 4 \%$ 
contribution to the total GT matrix element. 

Here, we consider two-body axial currents derived within the same $\chi$EFT
used to construct the NV2+3 interactions~\cite{Baroni:2018fdn}.  Moreover, we base our calculations 
on tree-level corrections only, and disregard the (problematic) N4LO loop contributions
discussed above. This choice is advantageous also because it allows for a
clearer comparison with the no-core shell model and coupled-cluster 
calculations of Ref.~\cite{Gysbers:2019uyb}, which are also based on tree-level 
axial currents alone ({\it albeit} derived in a $\Delta$-less $\chi$EFT).  Corrections at
N4LO in the present formulation are, in practice, subsumed
in the LECs of the theory, which have been determined by fits 
to experimental data.

 \begin{figure}
 \vspace*{.2in}
 \includegraphics[width=3.3in]{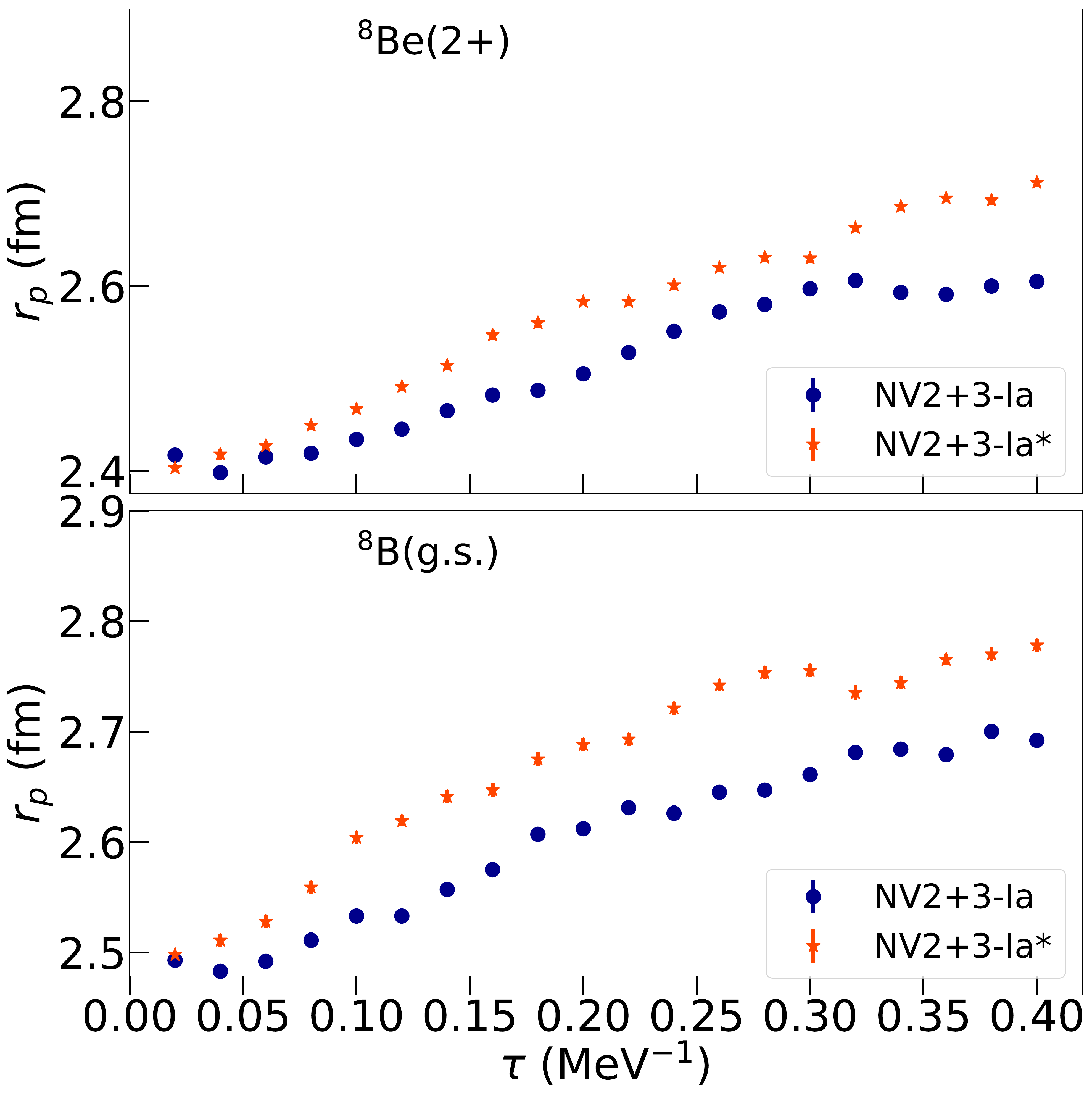}
 \caption{
 GFMC propagation of the point-proton radius of the first excited state of
 $^8$Be (upper panel) and $^8$B (lower panel) based on the NV2+3-Ia and NV2+3-Ia* Hamiltonians.}
\label{fig:rp}
\end{figure}

Before moving on to a (brief) discussion of these axial
currents, it is worthwhile pointing out that many-body corrections to
leading one-body transition operators have been shown to be crucial
for providing a quantitatively successful description of many nuclear electroweak
observables~\cite{Bacca:2014tla}, such as nuclear electromagnetic form 
factors~\cite{Piarulli:2012bn,Schiavilla:2018udt,NevoDinur:2018hdo,Marcucci:2015rca}, low-energy electroweak
transitions~\cite{Pastore:2009is,Girlanda:2010vm,Pastore:2011ip,Pastore:2012rp,Datar:2013pbd,Pastore:2014oda,Pastore:2017uwc},
and electroweak scattering~\cite{Pastore:2019urn}. They have also been used in 
studies of double beta decay matrix elements~\cite{Pastore:2017ofx,Cirigliano:2018hja,Cirigliano:2019vdj,Wang:2019hjy}.

The N3LO axial currents used in this work are represented diagrammatically Fig.~\ref{fig:currax}.
We refer the interested reader to Refs.~\cite{Baroni:2015uza,Baroni:2018fdn}
for additional details and explicit expressions of the operators; here, we only
note that we do not show diagrams that lead to vanishing contributions as well as
pion-pole terms which give negligible corrections to the 
matrix elements under study.

The LO term, which scales as $Q^{-3}$ in the power counting 
($Q$ denotes generically a low-momentum scale), is shown in panel (a)
and reads
\begin{equation}
{\bf j}^{\rm LO}_{5,a}({\bf q})=-\frac{g_A}{2}\, \tau_{i,a}\,  {\bm \sigma}_i\, {\rm e}^{i{\bf q}\cdot {\bf r}_i}\  ,
\end{equation}
where $g_A$ is the nucleon axial coupling constant
($g_A\,$=$\, 1.2723$~\cite{PDG}), $f_\pi$ is
the pion-decay amplitude ($f_\pi\,$=$\,92.4$ MeV), ${\bm \sigma}_i$ and ${\bm \tau}_i$
are the spin and isospin Pauli matrices of nucleon $i$, ${\bf q}$ is the external field momentum,
${\bf r}_i$ is the position of nucleon $i$, and the subscript $a$ specifies
the isospin component ($a\,$=$\,x,y,z$).

At N2LO there are two contributions (scaling as $Q^{-1}$).  The first one is a relativistic correction
to the single-nucleon operator at LO and is diagrammatically illustrated in panel (b), while
the second involves the excitation of a nucleon into a $\Delta$ by pion exchange, 
as illustrated in panel (c).  In the tables and figures below, we will denote these two contributions with 
N2LO-RC and N2LO-$\Delta$, respectively. We use the same notation introduced
in Ref.~\cite{Baroni:2018fdn} and write the cumulative N2LO contribution as
\begin{equation}
{\bf j}^{\rm N2LO}_{5,a}({\bf q})={\bf j}^{\rm N2LO}_{5,a}({\bf q};{\rm RC})
+{\bf j}^{\rm N2LO}_{5,a}({\bf q};\Delta) \ . 
\end{equation}
                                               
At N3LO (or $Q^{0}$ in the chiral expansion), there is a term of one-pion range
illustrated in panel (d), and a contact term shown in panel (e), which together
give the following N3LO correction
\begin{equation}
{\bf j}^{\rm N3LO}_{5,a}({\bf q})= {\bf j}^{\rm N3LO}_{5,a}({\bf q};{\rm OPE})
+{\bf j}^{\rm N3LO}_{5,a}({\bf q};{\rm CT})\ .
\end{equation}
We will denote the individual terms with N3LO-OPE 
and N3LO-CT, respectively. 

The configuration-space expressions of these currents are given 
in Eqs.~(2.7)--(2.10) of Ref.~\cite{Baroni:2018fdn}. Here, we limit ourselves
to report the expression of the N3LO contact term used in this work to explicitly 
show the relation between the LECs entering this axial current and the LEC $c_D$ in
the three-nucleon interaction. In $r$-space the 
N3LO-CT current reads
\begin{equation}
 {\bf j}^{\rm N3LO}_{5,a}({\bf q};{\rm CT})= z_0\, {\rm e}^{i\, {\bf q}\cdot {\bf R}_{ij}}\, \frac{ {\rm e}^{-z_{ij}^2}}{\pi^{3/2}}\,
\left({\bm \tau}_i\times{\bm \tau}_j\right)_a\,
\left({\bm \sigma}_i\times{\bm \sigma}_j\right)  \ ,
\label{eq:axct}
\end{equation}
where 
\begin{equation}
{\bf R}_{ij}=\left({\bf r}_i+{\bf r}_j\right)/2 \ ,  \qquad z_{ij}=r_{ij}/R_{\rm S} \ ,
\end{equation}
and $r_{ij}$ is the interparticle distance. 
The $\delta$-function in the contact axial current has been smeared by replacing it with a Gaussian cutoff 
of range $R_{S}$,  
\begin{equation}
\label{eq:ctsr}
	C_{R_{\rm S}}(r_{ij})=\frac{1}{\pi^{3/2}R_{\rm S}^3} e^{-(r_{ij}/R_{\rm S})^2}\, ,
\end{equation}
as previously done for the contact-like terms of the NV2 interactions. 
The adimensional LEC $z_0$ (reported in Table~\ref{tab:cdce}) is given by
\begin{eqnarray}
\label{eq:ez0}
z_0&=&\frac{g_A}{2}\,\frac{m_\pi^2}{f_\pi^2}\, \frac{1}{\left(m_\pi \,R_{\rm S}\right)^3}
\bigg[ -\frac{m_\pi}{4\, g_A\, \Lambda_\chi}\, c_D \nonumber\\
&&+\frac{m_\pi}{3} \left( c_3+ 2\, c_4 \right)
+\frac{m_\pi}{6\, m} \bigg] \  ,
\end{eqnarray}
where $c_D$ is the LEC multiplying one of the contact terms
in the three-nucleon interaction~\cite{Epelbaum:2002vt} given in Table~\ref{tab:cdce},
$\Lambda_\chi=1$ GeV is the chiral symmetry breaking scale, $c_3$ and $c_4$ are given in Table~\ref{tab:lec1},
and $m$ and $m_\pi$ are the average nucleon and pion masses. It has recently been
realized~\cite{Schiavilla:2017} that the relation between $z_0$ and $c_D$
had been given erroneously in the original reference~\cite{Gazit:2008ma}, a --
sign and a factor 1/4 were missing in the term proportional to $c_D$.

\begin{table}[bth]
\caption{
GFMC predictions for $A\leq 10$ nuclear states studied in this work,
compared to experimental values~\cite{Tilley:2002vg,Tilley:2004zz}. 
Numbers in parentheses are statistical errors for the GFMC calculations;
experimental errors, being negligible, are not indicated.  The dominant spatial symmetry (s.s.) of the nuclear wave function is given 
in the second column.}
\label{tb:energies}
\begin{ruledtabular}
\footnotesize
 \begin{tabular}{c c l l l }
   $^AZ(J^\pi;T)$                &
   \multicolumn{1}{c}{s.s.}      &
   \multicolumn{2}{c}{$E$ [MeV]} 
   \\ \cline{1-2} \cline{3-4}
   \\   
                            &
                            &
   \multicolumn{1}{c}{Ia{\color{white}*}}   &
   \multicolumn{1}{c}{Ia*}   &
   \multicolumn{1}{c}{Expt.}    
   \\
   \hline
   \\
   \nuc{6}{Li}$(1^+;0)$			    & [42]  & --31.97(6)  & --31.06(8)   & --31.99  
   \\
   				            
   \nuc{6}{He}$(0^+;0)$                     & [42]  & --29.32(4)  & --28.46(5)   & --29.27  
   \\
                                            
   \nuc{7}{Li}$(\frac{3}{2}^-;\frac{1}{2})$ & [43]  & --39.25(15) & --38.27(14)  & --39.25
   \\
                                            
   \nuc{7}{Li}$(\frac{1}{2}^-;\frac{1}{2})$ & [43]  &--39.18(15) &--37.66(15)    &--38.76 
   \\
 
   \nuc{7}{Be}$(\frac{3}{2}^-;\frac{1}{2})$ & [43]  & --37.75(8)  & --36.56(10)   & --37.60 
   \\
 
   \nuc{8}{He}$(0^+;2)$                     & [422] & --31.33(7)  & --28.53(6)    & --31.40
   \\
                                            
   \nuc{8}{Li}$(2^+;1)$                     & [431] & --41.59(10) & --38.89(7)    & --41.28 
   \\

   \nuc{8}{Li}$(1^+;1)$                     & [431] & --40.59(7)  & --37.78(7)    & --40.30 
   \\   
   
   \nuc{8}{B}$(2^+;1)$                      & [431] & --37.87(8)  & --35.63(8)    & --37.74  
   \\
                                            
   \nuc{8}{Be}$(2^+;0)$                     & [44]  & --54.07(7)  & --53.16(11)   & --53.47  
   \\
                                            
   \nuc{10}{B}$(1^+;0)$                     & [442] & --64.61(41) & --60.46(30)   &--64.03 
   \\
                                            
   \nuc{10}{C}$(0^+;1)$                     & [442]& --61.01(50)&    --56.65(22)  &--60.32 
   \\
   \end{tabular}
\end{ruledtabular}
\end{table}

\begin{table*}[tbh]
\begin{center}
\caption{
Gamow-Teller RMEs in $A\,$= 6, 7, 8, and 10 nuclei obtained with chiral
axial currents~\cite{Baroni:2018fdn} and VMC wave functions corresponding to the NV2+3-Ia/b and NV2+3-IIa/b (NV2+3-Ia/b* and NV2+3-IIa/b*) 
Hamiltonian models~\cite{Piarulli:2016vel,Piarulli:2014bda,Piarulli:2017dwd,Baroni:2018fdn}.
Columns labeled with LO, N2LO-(RC+$\Delta$), N3LO-OPE, and N3LO-CT refer to the contributions given by the diagrams
illustrated in panel (a), panels (b) plus (c), panel (d), and panel (e) of Fig.~\ref{fig:currax}, respectively.
The cumulative results are reported in the column labeled ``Total'', while results including only corrections beyond LO are
listed under ``Total--LO''. Experimental values from Refs.~\cite{Knecht:2012fs,Suzuki:2003,Chou:1993zz,1986WA01,1989BA31}
are given in the last column. The dominant spatial symmetries of the VMC wave functions are reported in the first column.
Statistical errors associated with the Monte Carlo integrations are not shown, but are below 1\%.
}
\label{tab:break}
\footnotesize
\begin{tabular}{c l l l l l l l l l }
\hline
\hline
\\
Transition & Model 
                     & \multicolumn{1}{l}{LO}
                     & \multicolumn{1}{l}{N2LO-(RC+$\Delta$)} 
                     & \multicolumn{1}{l}{N3LO-OPE} 
                     &\multicolumn{1}{l}{N3LO-CT}
                     & \multicolumn{1}{l}{Total--LO}      
                     &\multicolumn{1}{l}{Total}
                     & \multicolumn{1}{l}{Expt.} \\
\hline \\ [2pt]
{\footnotesize$^6$He($0^+$;$1$)$\rightarrow$ $^6$Li($1^+$;$0$)} & Ia{\color{white}I} (Ib)   &2.200 (2.254)     &0.022 (0.056) & 0.039 (0.064)&--0.005 (--0.068) &0.056 (0.052) & 2.256 (2.306) &    2.1609(40)\\
         {\footnotesize   [42]$\rightarrow$[42]  }              & IIa (IIb)                 &2.207 (2.212)     &0.027 (0.043) &0.043 (0.055) &--0.034 (--0.082) &0.036 (0.016) &2.243 (2.228)  & \\
                                                                & Ia*{\color{white}I} (Ib*) &2.192 (2.256)     &0.021 (0.056) & 0.038 (0.063)&--0.054 (--0.097) &0.005 (0.022) & 2.197 (2.279) & \\
                                                                & IIa* (IIb*)               &2.202 (2.218)     &0.027 (0.044) &0.043 (0.056) &--0.057 (--0.090) &0.014 (0.010) &2.216 (2.228)  & \\ [8pt]
{\footnotesize$^7$Be($\frac{3}{2}^-$;$\frac{1}{2}$)$\rightarrow ^7$Li($\frac{3}{2}^-$;$\frac{1}{2}$)}  & Ia{\color{white}I} (Ib)  &2.317 (2.294) &0.099 (0.162)& 0.076 (0.118)&--0.010 (--0.148) &0.165 (0.133) &2.482 (2.427) & 2.3556(47)\\
         {\footnotesize [43]$\rightarrow$[43]}                                                         & IIa (IIb)                &2.293 (2.309) &0.102 (0.153)& 0.078 (0.113)&--0.070 (--0.190) &0.110 (0.076)&	2.403 (2.385)\\
			                                                                               & Ia*{\color{white}I} (Ib*)&2.327 (2.307) &0.098 (0.161)& 0.076 (0.117)& --0.121 (--0.212)&0.053 (0.066) &2.380 (2.373) & \\
                                                                                                       & IIa* (IIb*)              &2.296 (2.316) &0.103 (0.154)& 0.078 (0.114)&--0.120 (--0.210) &0.061 (0.058) &2.357 (2.374)&\\ [8pt]
{\footnotesize$^7$Be($\frac{3}{2}^-$;$\frac{1}{2}$)$\rightarrow ^7$Li($\frac{1}{2}^-$;$\frac{1}{2}$)}  & Ia{\color{white}I} (Ib)  &2.157 (2.119) &0.066 (0.122)& 0.063 (0.100)&	--0.009 (--0.125)&0.121 (0.096)&	2.278 (2.215)&  2.1116(57)    \\				
         {\footnotesize [43]$\rightarrow$[43]}                                                         & IIa (IIb)                &2.128 (2.145) &0.069 (0.111)& 0.065 (0.095) &--0.059 (--0.162)&0.074 (0.044)&	2.202 (2.189)&\\
                                                                                                       & Ia*{\color{white}I} (Ib*)&2.158 (2.124) &0.065 (0.119)& 0.063 (0.099)&	--0.103 (--0.180)&0.025 (0.038)&	2.183 (2.162) & \\				
                                                                                                       & IIa* (IIb*)              &2.131 (2.148) &0.067 (0.111)& 0.064 (0.095)&	--0.101 (--0.178)&0.030 (0.028)&	2.161 (2.176) &\\[8pt]
{\footnotesize$^8$Li($2^+$;1)$\rightarrow ^8$Be($2^+$;$0$)}  & Ia{\color{white}I} (Ib)&  0.147 (0.092)&	0.032 (0.028)&	0.011 (0.011)&	--0.001 (--0.014)& 0.041 (0.031)&	0.188 (0.123) & 0.284 Ref.~\cite{1986WA01}\\
         {\footnotesize [431]$\rightarrow$[44]}              & IIa (IIb)              &  0.144 (0.101)&	0.031 (0.033)&	0.010 (0.011)&	--0.008 (--0.019)& 0.033 (0.025)&	0.177 (0.126) & 0.190 Ref.~\cite{1989BA31}\\ 
                                                             & Ia*{\color{white}I} (Ib*)& 0.148 (0.099)&0.032 (0.033)&	0.010 (0.012)&	--0.016 (--0.020)& 0.026 (0.025)&	0.174 (0.124) & \\
                                                             & IIa* (IIb*)              & 0.124 (0.121)&0.032 (0.037)&	0.010 (0.013) &	--0.014 (--0.023)& 0.028 (0.027)&	0.152 (0.148) &\\ [8pt]
{\footnotesize$^8$B($2^+$;1)$\rightarrow ^8$Be($2^+$;$0$)}   & Ia{\color{white}I} (Ib)  & 0.146 (0.092)&0.032 (0.032) &0.011 (0.011) &--0.001 (--0.014) &0.042 (0.030) &0.188 (0.122)& 0.269(20) \\
         {\footnotesize [431]$\rightarrow$[44]}              & IIa (IIb)                & 0.144 (0.102)&0.031 (0.033) &0.010 (0.011) &--0.008 (--0.019) &0.033 (0.026) &0.177 (0.128) &\\   
                                                             & Ia*{\color{white}I} (Ib*)& 0.148 (0.098)&0.032 (0.034) &0.010 (0.012) &--0.016 (--0.020) &0.026 (0.025) &0.174 (0.123)& \\
                                                             & IIa* (IIb*)              & 0.126 (0.118)&0.032 (0.037) &0.010 (0.013) &--0.014 (--0.022) &0.028 (0.027) &0.154 (0.145) &\\ [8pt]
{\footnotesize$^8$He($0^+$;2)$\rightarrow ^8$Li($1^+$;$1$)}  & Ia{\color{white}I} (Ib)     & 0.386 (0.363) & 0.030 (0.034) & 0.009 (0.012) & --0.001 (--0.014) & 0.038 (0.032) & 0.424 (0.396) & 0.512(6)\\
        {\footnotesize [422]$\rightarrow$[431]}              & IIa (IIb)                   & 0.465 (0.370) & 0.032 (0.034) & 0.012 (0.011) & --0.009 (--0.017) & 0.035 (0.028) & 0.500 (0.398) & \\   
                                                             & Ia*{\color{white}I} (Ib*)   & 0.362 (0.377) & 0.031 (0.034) & 0.009 (0.014) & --0.010 (--0.022) & 0.029 (0.026) & 0.391 (0.402) &  \\
                                                             & IIa* (IIb*)                 & 0.481 (0.364) & 0.033 (0.035) & 0.012 (0.012) & --0.017 (--0.019) & 0.029 (0.028) & 0.510 (0.391) & \\ [8pt]  
{\footnotesize$^{10}$C($0^+$;1)$\rightarrow ^{10}$B($1^+$;$0$)} & Ia{\color{white}I} (Ib)  & 1.940 (2.118) & 0.003 (0.026) & 0.044 (0.068) & --0.006 (--0.081) &0.041 (0.011)   &1.981 (2.129) & 1.8331(34) \\
        {\footnotesize [442]$\rightarrow$[442]}                 & IIa (IIb)                & 2.157 (2.176) &-0.002 (0.026) & 0.046 (0.072) & --0.042 (--0.120) &0.002 (--0.022)  &2.159 (2.154) & \\
                                                                & Ia*{\color{white}I} (Ib*)&  2.015 (2.123) &  0.024 (0.022) & 0.059 (0.069) &  --0.119 (--0.117) & --0.037 (--0.028) & 1.978 (2.095) & \\
                                                                & IIa* (IIb*)              & 2.071 (2.143) &0.015 (0.031) & 0.061 (0.072) & --0.126 (--0.123) &--0.050 (--0.020) &2.021 (2.123) &\\ [8pt]
\hline
\end{tabular}
\end{center}
\end{table*}

\begin{center}
\begin{table*}[tbh]
\small\addtolength{\tabcolsep}{4.6pt}
\caption{
Gamow-Teller RMEs in $A\,$= 6, 7, 8, and 10 nuclei obtained with chiral
axial currents~\cite{Baroni:2018fdn} and GFMC (VMC) wave functions corresponding to the NV2+3-Ia and NV2+3-Ia* 
Hamiltonian models~\cite{Piarulli:2016vel,Piarulli:2014bda,Piarulli:2017dwd,Baroni:2018fdn}.
Results corresponding to the one-body current at LO (column labeled ``LO''), and to the sum of all the corrections beyond
LO (column labeled ``Total--LO'') are given, along with the cumulative contributions (column labeled `Total') 
to be compared with the experimental data~\cite{Knecht:2012fs,Suzuki:2003,Chou:1993zz,1986WA01,1989BA31} reported in the last row.
Results from Ref.~\cite{Pastore:2017uwc} based on the AV18 and IL7 nuclear Hamiltonian are also shown
where available. 
Statistical errors associated with the Monte Carlo integrations are not shown, but are  below 1\%.
Transitions to $^8$Be are affected by an additional systematic
error of $\approx 5\%$, see text for explanation.
}
\label{tab:gfmcgt}
\begin{tabular}{c c c l l l l }
\hline
\hline
\\[1pt]
Transition & Model &
   \multicolumn{1}{c}{s.s.} &
   \multicolumn{1}{l}{LO} &
   \multicolumn{1}{l}{Total--LO} &
   \multicolumn{1}{l}{Total} &
   \multicolumn{1}{l}{Expt.}  \\ [1pt]
\hline
\\[1pt]
{\footnotesize$^6$He($0^+$;$1$)$\rightarrow$ $^6$Li($1^+$;$0$)}
& Ia{\color{white}I*}  & [42]$\rightarrow$[42]   &2.130(2.200)   & 0.070(0.056)   & 2.201(2.256)  &  2.1609(40)\\
& Ia*{\color{white}I}  &                         &2.107(2.192)   & 0.011(0.005)   & 2.118(2.197)  & \\ 
& Ref.~\cite{Pastore:2017uwc} &                  &2.168(2.174)   & 0.037(0.030)   & 2.205(2.211)  & \\ 
[2pt]
\hline
\\[1pt]
{\footnotesize$^7$Be($\frac{3}{2}^-$;$\frac{1}{2}$)$\rightarrow ^7$Li($\frac{3}{2}^-$;$\frac{1}{2}$)}
& Ia{\color{white}I*}         & [43]$\rightarrow$[43] &  2.273(2.317)     & 0.164(0.165)     &  2.440(2.482) &  2.3556(47)\\
& Ia*{\color{white}I}         &                       &  2.286(2.327)     & 0.052(0.053)     &  2.338(2.380) & \\
& Ref.~\cite{Pastore:2017uwc} &                       &  2.294(2.334)     & 0.061(0.050)     &  2.355(2.384) & \\ 
[2pt]
\hline
\\[1pt]
{\footnotesize$^7$Be($\frac{3}{2}^-$;$\frac{1}{2}$)$\rightarrow ^7$Li($\frac{1}{2}^-$;$\frac{1}{2}$)}
& Ia{\color{white}I*}         & [43]$\rightarrow$[43]   & 2.065(2.157)  & 0.103(0.121)     &  2.168(2.278)   & 2.1116(57) \\
& Ia*{\color{white}I}         &                         & 2.061(2.158)  & 0.009(0.025)     &  2.070(2.183)   & \\
& Ref.~\cite{Pastore:2017uwc} &                         &  2.083(2.150) & 0.046(0.046)     &  2.129(2.196)   & \\ 
[2pt]
\hline
\\[1pt]
{\footnotesize$^8$Li($2^+$;1)$\rightarrow ^8$Be($2^+$;$0$)}
& Ia{\color{white}I*}       & [431]$\rightarrow$[44]  &0.074(0.147)    &0.029(0.041)  & 0.103(0.188) & 0.284~Ref.~\cite{1986WA01} \\
& Ia*{\color{white}I}       &                         &0.096(0.148)    &0.025(0.026)  & 0.120(0.174) & 0.190~Ref.~\cite{1989BA31} \\ [2pt]
\hline
\\[1pt]
{\footnotesize$^8$B($2^+$;1)$\rightarrow ^8$Be($2^+$;$0$)}
& Ia{\color{white}I*}       & [431]$\rightarrow$[44]  &0.091(0.146)  & 0.035(0.042)  &0.125(0.188) & 0.269(20) \\
& Ia*{\color{white}I}       &                         &0.102(0.148)  & 0.024(0.026)  &0.126(0.174) & \\[2pt]
\hline
\\[1pt]
{\footnotesize$^8$He($0^+$;2)$\rightarrow ^8$Li($1^+$;$1$)}
& Ia{\color{white}I*}       & [422]$\rightarrow$[431]  &0.262(0.386) &0.040(0.038) &0.302(0.424) & 0.512(6) \\
& Ia*{\color{white}I}       &                          &0.297(0.362) &0.025(0.029) &0.322(0.391) & \\[2pt]
\hline
\\[1pt]
{\footnotesize$^{10}$C($0^+$;1)$\rightarrow ^{10}$B($1^+$;$0$)}
& Ia{\color{white}I*}       & [442]$\rightarrow$[442]  & 1.928(1.940)  &0.050(0.041)      &1.978(1.981)  & 1.8331(34) \\
& Ia*{\color{white}I}       &                          & 2.086(2.015)  &--0.031(--0.037)  &2.055(1.978)  & \\ 
& Ref.~\cite{Pastore:2017uwc} &                        &  2.032(2.062) & 0.016(0.015)     &2.048(2.077)  & \\
[2pt]
\hline
\hline
\end{tabular}
\end{table*}
\end{center}
\section {Results}
\label{sec:res}

The GFMC energies of the nuclei of interest calculated using the  
NV2+3-Ia and NV2+3-Ia* models are listed in Table~\ref{tb:energies}
along with the dominant spatial symmetry (s.s.) of the variational 
wave functions~\cite{PhysRevC.73.034317}.
The energies are obtained using $\approx 80,000$ walkers, and are 
all well converged by 30 unconstrained steps~\cite{WPCP00}. All the GFMC
results presented in this article (but for the two cases discussed below) 
are averages over the imaginary time $\tau$ from 0.2 to 0.82 MeV$^{-1}$. 
Results obtained with the NV2+3-Ia interaction are in statistical agreement 
with those published in Ref.~\cite{Piarulli:2017dwd} based on the 
same nuclear Hamiltonian. Model NV2+3-Ia leads to predictions that are in excellent 
agreement with the data. We also report for the first time results based on the
second generation of NV2+3 interactions, specifically model NV2+3-Ia*, whose
three-nucleon interaction has been constrained by fits to the
experimental trinucleon binding energies
and tritium GT matrix element. Results obtained with the NV2+3-Ia$^*$
Hamiltonian display a somewhat less satisfactory agreement with the 
experimental data, but still less than 4\% away from them.

A typical imaginary-time evolution of the GFMC transition matrix elements
is shown in Fig.~\ref{fig:it}.
As can be seen, there is a rapid drop of 3\% from the initial VMC estimate
at $\tau\,$=$\,0$ that reaches a stable value around 0.2 MeV$^{-1}$.  The results for all
transitions presented in this article are averages over $\tau$ from 0.2 to 0.82 MeV$^{-1}$, 
as indicated by the dashed lines, with statistical errors denoted by the
solid lines. The calculations of weak transitions involving the $(J^\pi,T)=(2^+,0)$ 
state of $^8$Be and the ground state~of $^8$B
are treated differently. For these two states,
we observe that the binding energy, magnitude of the quadrupole moment, and point-proton 
radius all increase monotonically as the imaginary time increases.  This can be appreciated
in Fig.~\ref{fig:rp} where we show the point-proton radii of the two nuclear states.
We interpret this behavior as an indication that the resonant excited state of $^8$Be 
is dissolving into two separated $\alpha$'s, while $^8$B is breaking into $p+^7$Be. 
In the case of $^8$Be, this issue has been addressed already in Refs.~\cite{Wiringa:2000gb,Pastore:2014oda,Datar:2013pbd}.
Here, we use similar techniques to treat these systems and extract matrix elements
from the GFMC data. In particular, we note that the ground-state energy of $^8$Be 
drops very quickly as the imaginary time increases and reaches stability
around $\tau\sim 0.1$ MeV$^{-1}$. The transition matrix elements 
involving the two dissolving states have been determined 
assuming that, also for these states, $\tau \approx 0.1$ MeV$^{-1}$ is the point 
at which  spurious contaminations in the nuclear
wave functions have been removed by the
GFMC propagation. We then average in a small interval around this point, typically
between 0.06 and 0.14 MeV$^{-1}$. Calculations involving these two systems
are  clearly affected by a systematic error. To have a rather rough estimate of this
error, we study the sensitivity of the extracted matrix elements
with respect to variations in the imaginary time interval selected for the 
averaging. We find that such a procedure generates an additional uncertainty of $\approx 5 \%$ 
which is added in quadrature to the statistical one, and quoted in Table~\ref{tab:gfmcgt} below of GFMC results.

\begin{figure}%
\includegraphics[height=1.8in]{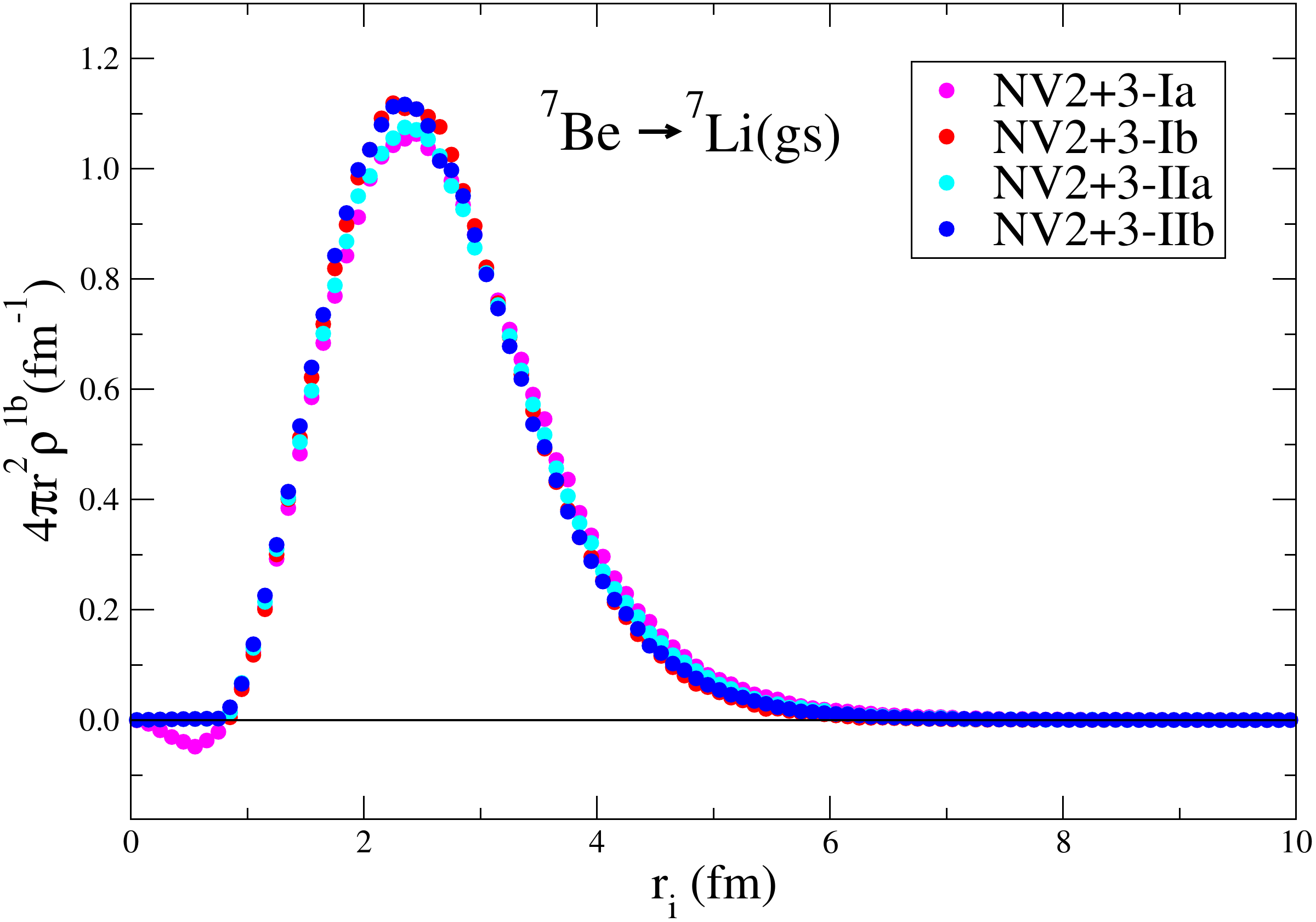}
\caption{(Color online) One-body density--defined in  Eq.~(\ref{eq:rho1})--of the $^7$Be to $^7$Li(gs)
GT RME obtained with models NV2+3-Ia/b and NV2+3-IIa/b.(see text for explanation). }
\label{fig:1density7}
\end{figure}
\begin{figure}%
\includegraphics[height=1.8in]{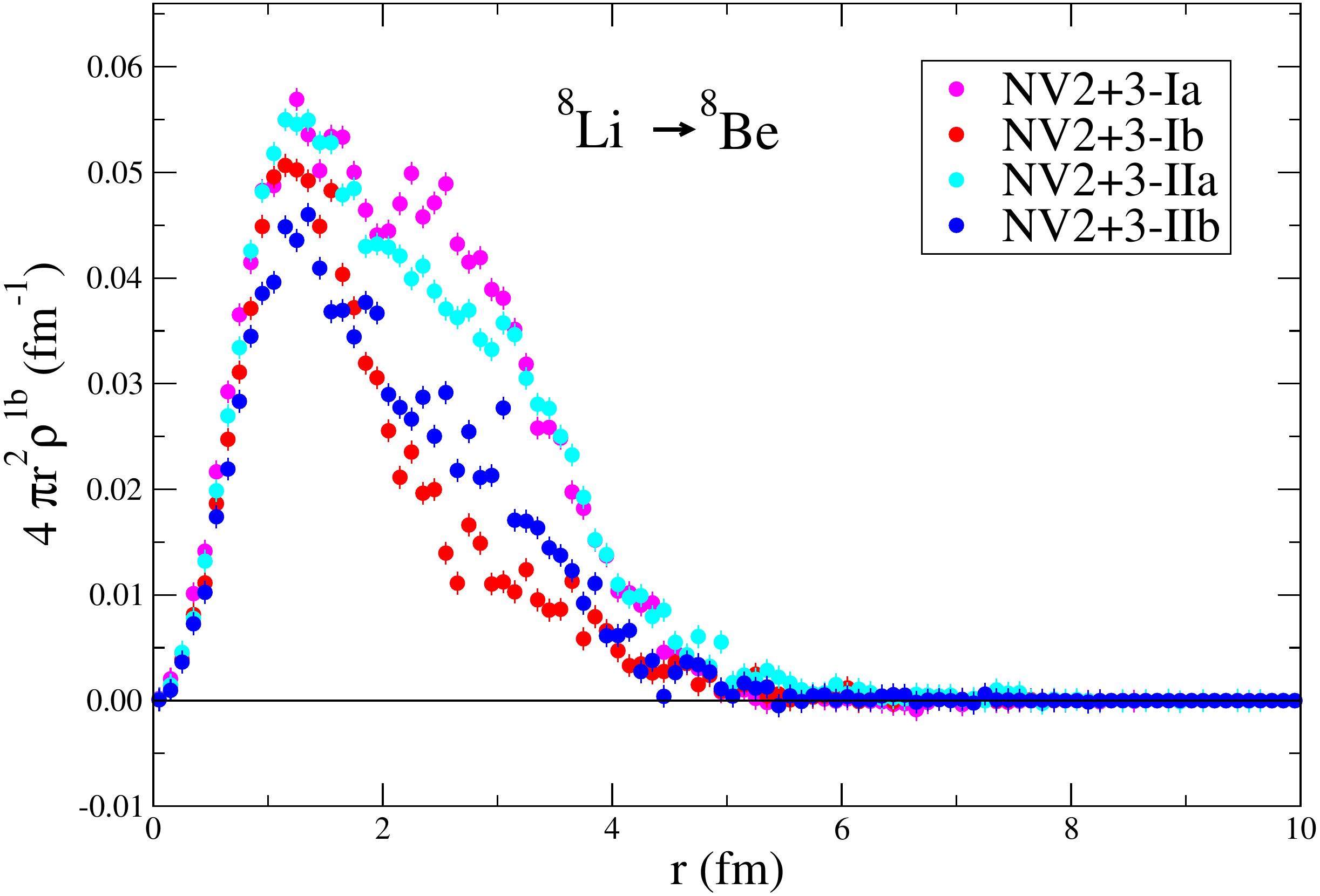}
\caption{(Color online) Same as Fig.~\ref{fig:1density7} but for the $^8$Li to $^8$Be
GT RME. }
\label{fig:1density8}
\end{figure}
 
\subsection {Weak Transitions in $A$=6--10 Nuclei}
\label{subsec:matrix}

In this section, we present results for the GT
reduced matrix element (RME) defined as
\begin{equation}
{\rm RME}=\frac{\sqrt{2\, J_f+1}}{g_A} \,
\frac{\langle J_f M | j^z_{5,\pm} | J_i M\rangle}{\langle J_iM, 10 |J_f M\rangle} \ ,
\end{equation}
where  $j^z_{5,\pm}$ is the $z$-component (at vanishing momentum transfer) of the
charge-raising/lowering current with ${\bf j}_{5,\pm} \,$=$\, {\bf j}_{5,x}\pm i\, {\bf j}_{5,y}$,
and $\langle J_iM, 10 |J_f M\rangle$ is a Clebsch-Gordan coefficient. 

Results for the GT RMEs in $A\,$=$\,6$--10
nuclei based on variational wave functions are reported in Table~\ref{tab:break}
for the eight different NV2+3 interactions discussed above, 
namely the NV2+3-Ia/b, NV2+3-IIa/b and corresponding starred models.
The one-body axial current at LO, illustrated in panel (a) of Fig.~\ref{fig:currax},
leads to contributions to the matrix elements reported in the third column of 
Table~\ref{tab:break}. One-body relativistic corrections (N2LO-RC) and two-body currents of one-pion
range (N2LO-$\Delta$) at N2LO, displayed in panels (b) and (c) of Fig.~\ref{fig:currax}, 
are added up and given in the fourth column of Table~\ref{tab:break} labeled with N2LO-(RC+$\Delta$). 
A rough estimate of the size of the RC corrections can be 
obtained suppressing each LO term by a factor of $(Q/m)^2\approx 0.01$, where we used 
a ``typical'' nucleon's low-momentum $Q\approx 100$ MeV. 
Contributions at N3LO are given in the columns labeled
by N3LO-OPE and N3LO-CT, corresponding
to the one-pion range and contact currents  
(and displayed in panels (d) and (e) of Fig.~\ref{fig:currax}). 
The cumulative contributions are given in the next to last column, while the
contributions beyond LO only in the column labeled ``Total--LO''.
Experimental data from Refs.~\cite{Knecht:2012fs,Suzuki:2003,Chou:1993zz,1986WA01,1989BA31}
are reported in the last column of Table~\ref{tab:break}. 

All the calculations use axial currents at tree-level which are
consistent with the specific NV2+3 model used to generate
the VMC wave functions.  VMC results based on different nuclear Hamiltonians
are qualitatively in agreement. In particular, for the $A=3,6,7$ and $10$  systems
the LO contribution provides about $97 \%$ of the total matrix elements 
with currents beyond LO
giving the remaining $\lesssim 3\%$ correction.   This correction adds up constructively to the LO contribution
for all nuclei being considered,
but for the $A\,$=$\,10$ transition.   For this last transition,
we find that the contributions beyond LO give a correction
that quenches the LO results obtained with all the starred models,
and with the un-starred NV2+3-IIb interaction. More details about this calculation 
will be given in the following section. 
We emphasize that the ``Total--LO'' column includes, in addition to two-body 
contributions, also a small correction resulting from the
one-body N2LO-RC current. 

Transitions involving $A\,$=$\,8$ nuclei exhibit a large suppression at LO. 
This behavior is attributable to the fact that the initial and
final VMC wave functions are characterized by different dominant spatial
symmetries, which make their overlap small compared to cases in which 
both the initial and final states display the same dominant spatial symmetry. 
As a consequence, in these cases the LO term is only about $\approx \,$40--50\% of the 
total matrix element, with two-body currents providing a large correction. 
Two-body currents, while improving the agreement with the experimental 
values, are insufficient to fully explain them.  Because of the 
reduced overlap between dominant components in the wave functions,
these matrix elements are particularly sensitive to small components,
which are poorly constrained and model dependent. This can be appreciated 
by looking at the one-body transition densities, $\rho^{\rm 1b}(r_i) $, defined as 
\begin{equation}
\label{eq:rho1}
 {\rm RME}({\rm 1b})={\rm RME}({\rm LO}) = 4\, \pi \int dr_i\, r_i^2\, \rho^{\rm 1b}(r_i)   \, ,
\end{equation}
where $r_i$ is the distance of nucleon $i$ from the center-of-mass of 
the system. 

In Figs.~\ref{fig:1density7} and~\ref{fig:1density8}, we show one-body
densities for two transitions, namely the $\epsilon$-capture of the 
$^7$Be ground state to the $^7$Li ground state and the $^8$Li $\beta$-decay. 
The former, involves initial and final states with the same [43] (dominant)
spatial symmetry, while for the latter the initial state is in a [431] spatial symmetry 
configuration and the final state is in a [44] one. 
The densities are calculated using the NV2+3-Ia/b and NV2+3-IIa/b
interactions. From the figures we can see that the $A\,$=$\,7$ one-body
densities are well constrained and essentially model independent, 
while the $A\,$=$\,8$ ones are particularly sensitive
to the nuclear Hamiltonian used to generate the wave functions.
Of course, these considerations are based on VMC results.  A GFMC
propagation might mitigate the observed model dependence.

\begin{figure}%
\includegraphics[height=2.5in]{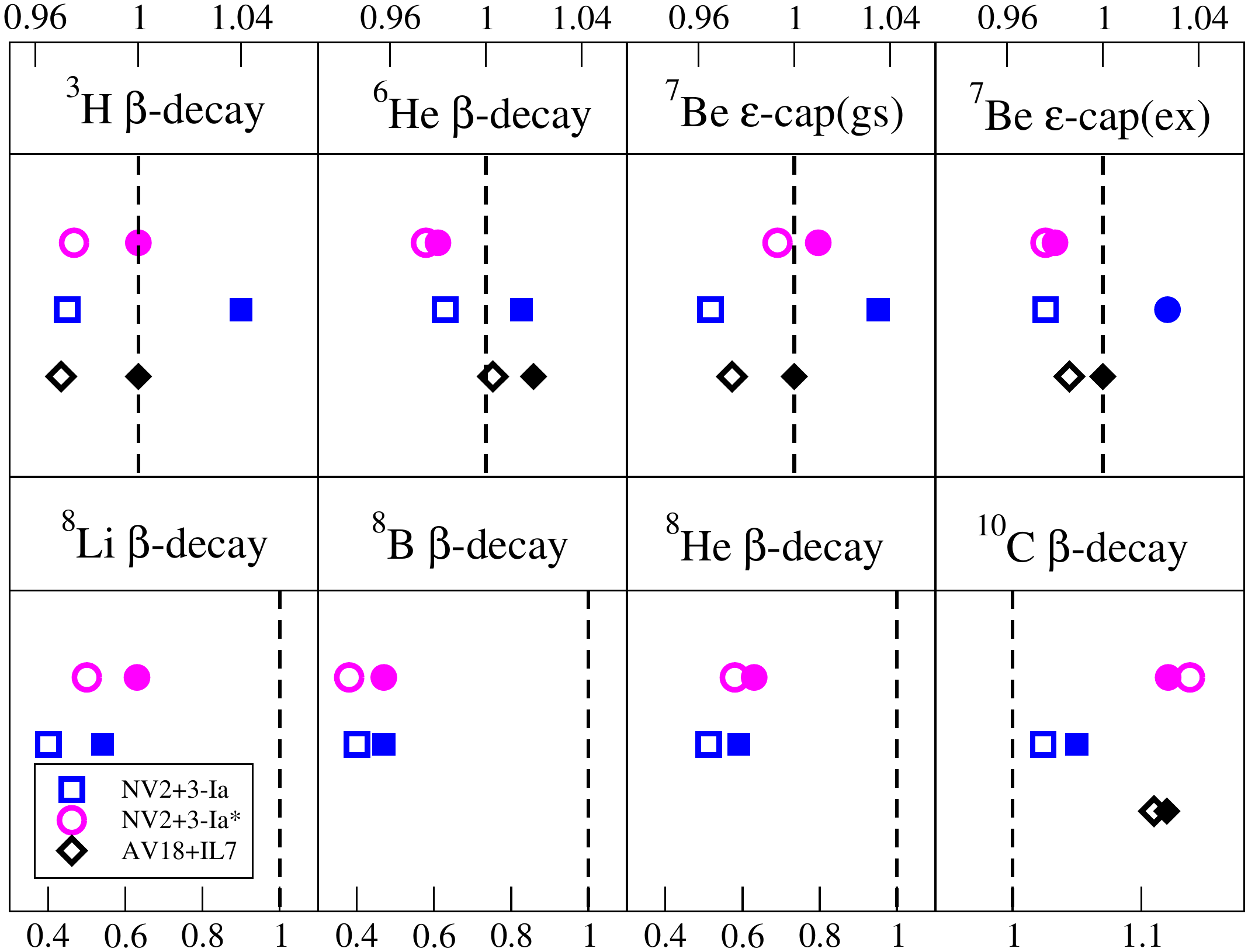}
\caption{
 (Color online) Ratios of GFMC to experimental values
 of the GT RMEs in the $^3$H, $^6$He, $^7$Be, $^8$B, $^8$Be, $^8$He and $^{10}$C weak transitions.  Theory
 predictions correspond to the $\chi$EFT axial current at LO (empty symbols) and up to 
 N3LO (filled symbols) obtained with the NV2+3-Ia and NV2+3-Ia* models. 
 Results from Ref.~\cite{Pastore:2017uwc} based on the AV18+IL7 nuclear Hamiltonian
 and N4LO currents from Ref.~\cite{Baroni:2015uza} are also shown. Results for the $^3$H weak transition were 
 reported in Ref.~\cite{Baroni:2018fdn}.
 }
\label{fig:rmes}
\end{figure}

In all the cases we studied, the contact-like current at N3LO provides 
a correction that quenches the LO terms, while the currents
of one-pion range at N2LO and N3LO add up constructively to the 
LO contributions (see Table~\ref{tab:break}). The main difference
between the un-starred and starred calculations is observed in the size 
of the contact contribution. In particular, starred models are characterized
by a larger value of $z_0$ (see Table~\ref{tab:cdce}), which in 
turns leads to a larger (in magnitude) N3LO-CT correction. 

We performed GFMC propagations only for the NV2+3-Ia and NV2+3-Ia* models. 
GFMC results are reported
in Table~\ref{tab:gfmcgt}, where, for completeness, we also show
the corresponding VMC values in parentheses along with the GFMC
results from Ref.~\cite{Pastore:2017uwc}.  We summarize
the GFMC results in  Fig.~\ref{fig:rmes} and compare them (where possible) to the
results of Ref.~\cite{Pastore:2017uwc} based on the AV18+IL7
nuclear Hamiltonian.

The effect of the GFMC 
propagation in imaginary time is to reduce the VMC results by 
$\lesssim 4\%$ in all selected transitions (but for the $A\,$=$\,10$
transition obtained with the NV2+3-Ia* model). The agreement with 
the data, after the inclusion of two-body currents, is at 
the $\approx 2\%$ ($\lesssim 2\%$) level for the $A\,$=$\,6$ transition
with the NV2+3-Ia (NV2+3-Ia*) model; and 
at the $\lesssim 4\%$ ($\lesssim 1\%$) level for the $A\,$=$\,7$ cases 
with the NV2+3-Ia (NV2+3-Ia*) model. These results are in 
agreement with those obtained for the same transitions in the
calculations of Ref.~\cite{Pastore:2017uwc} which were based on 
the AV18+IL7 interactions.
The NV2+3 models lead to a more satisfactory agreement with the 
data for the $A\,$=$\,6$ RME primarily because, with these interactions,
the LO term is $2\%$ smaller than obtained using AV18-IL7 model.

The largest discrepancy generated by the use of different
nuclear Hamiltonians, including AV18+IL7,
is observed in the $A\,$=$\,10$ transition. This can be appreciated
looking at both Table~\ref{tab:break} and Table~\ref{tab:gfmcgt}.
From the former, we observe a rather large cutoff dependence 
(models a vs models b), and also a large sensitivity to the class
(either I or II) used to generate the nuclear wave functions. 
From Table~\ref{tab:gfmcgt}, we see that the results of Ref.~\cite{Pastore:2017uwc}, 
based on the AV18+IL7 Hamiltonian, lie between models Ia and Ia*.
This large model and cutoff dependence can be traced back to the existence of two nearby $J^\pi \,$=$\, 1^+$ 
excited states in $^{10}$B, the lower one a predominantly $^3$S$_1$[442] state and the upper 
one a $^3$D$_1$[442] state (in $LS$ coupling), which are only 1 MeV apart.
The transition from the $^{10}$C(0$^+$) state, which is predominantly $^1$S$_0$[442], is 
large in the $S \rightarrow S$ components, but about five times smaller in the $S \rightarrow D$ components.
This makes the GT matrix element particularly sensitive to the exact mixing of the $^3$S$_1$ and $^3$D$_1$ 
components in the two $^{10}$B(1$^+$) states produced by a given Hamiltonian.
It would appear that none of the interactions models studied here gets quite the right mixing of these components.
In particular, results based on the NV2+3-Ia and NV2+3-Ia* interactions over-predict the data by $\sim7\%$ and $12\%$, 
respectively, which gives an indication of the spread of the theoretical estimates.

Predictions for the RMEs of $A\,$=$\,8$ transitions 
are the first QMC calculations for these systems that include 
corrections from two-body axial currents.
As discussed above, RMEs are suppressed at leading order which gives 
only $\approx 50 \%$ and $\approx 40 \%$ of the experimental values 
for the $^8$B$\,\rightarrow\,^8$Be and $^8$He$\,\rightarrow\,^8$Li transitions.
Two-body currents provide about 20--30\% correction in the right 
direction which is, however, still insufficient to reach agreement with 
the experimental data. These transitions are challenging not only from 
the theoretical but also from the experimental point of view.
For example, 
in Tables~\ref{tab:break} and~\ref{tab:gfmcgt}  we quote
two results for the  RME of the $^8$Li$\, \rightarrow\,^8$Be decay
obtained from the log($ft$)-values of Ref.~\cite{1986WA01}
and Ref.~\cite{1989BA31} via
the following formula~\cite{Chou:1993zz}
\begin{equation}
{\rm RME}({\rm EXPT}) = \frac{1}{g_A}\sqrt{2J_i+1}\sqrt{\frac{6139\pm7}{ft}} \ ,
\end{equation}
where $J_i$ is the angular momentum of the parent nucleus. The Fermi transition
strength is small enough in this case that it can be neglected in the above formula.
We then obtain two values, namely, RME(Ref.~\cite{1986WA01})=0.284, and  
RME(Ref.~\cite{1989BA31})=0.190. We note that Refs.~\cite{Suzuki:2003,Hardy:2014qxa},
report a different overall factor of $6147$ instead of $6139$ in 
the formula given above. In our estimate we used 
$g_A\,$=$\, 1.2723$. Despite this additional uncertainty in the 
deduced experimental values, our predictions
still severely underestimate the data. For example, 
the calculated Ia* RME provides only $\approx 40 \%$ and
$\approx 60 \%$ of the experimental values given in Ref.~\cite{1986WA01}
and Ref.~\cite{1989BA31}, respectively.

\begin{figure*}%
\includegraphics[height=4.4in]{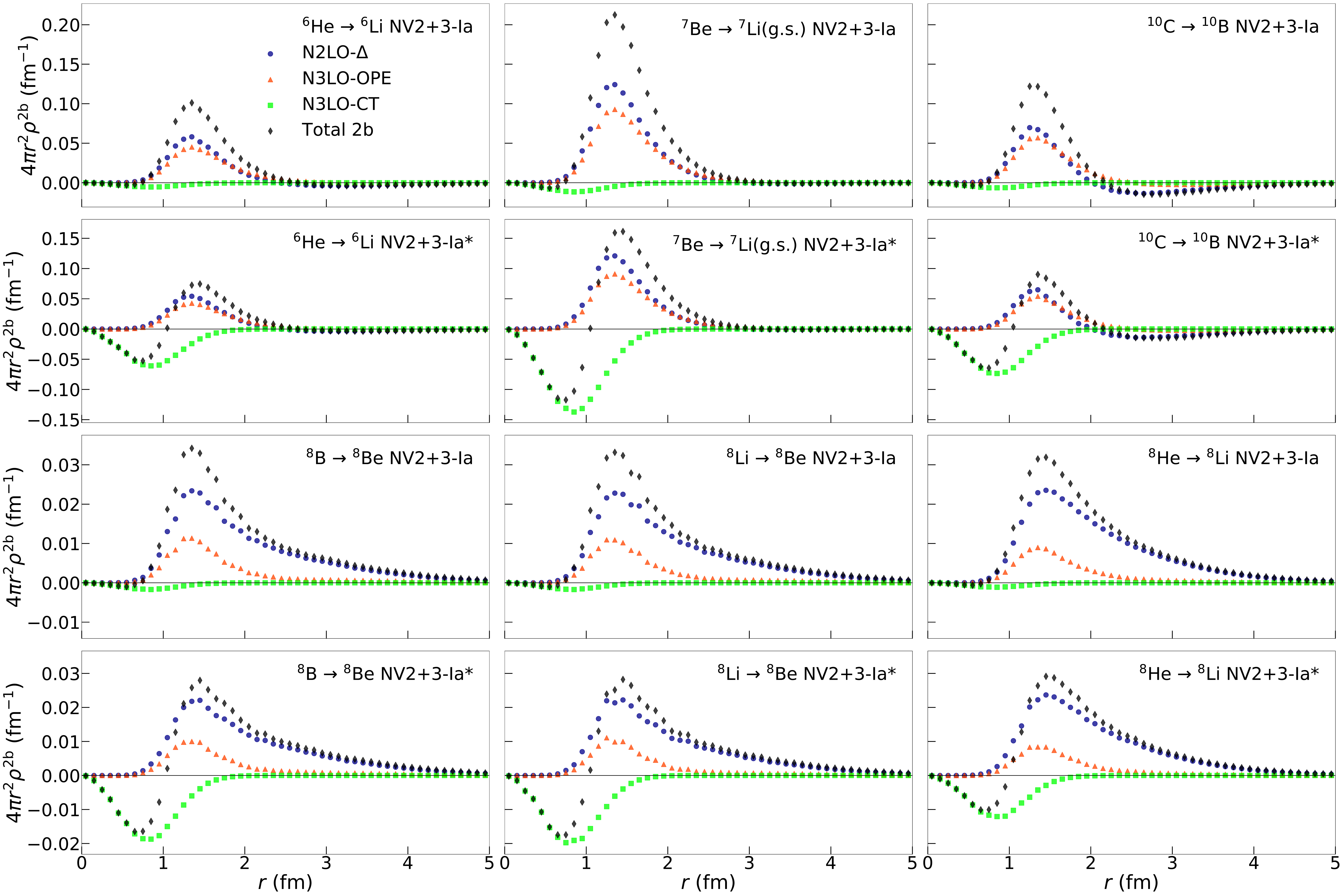}
\caption{(Color online)   Two-body transition density---see Eq.~(\ref{eq:rho2})---for 
selected nuclei obtained with the NV2+3-Ia and NV2+3-Ia* models (see text for explanation). }
\label{fig:2panel}
\end{figure*}

\subsection {Two--Body Transition Densities}
\label{subsec:trans}

In order to have a better understanding of the two-body terms in the axial current
and their contributions, it is helpful to study the associated two-body transition densities,
which we define as~\cite{Schiavilla:1998je}
\begin{equation}
\label{eq:rho2}
 {\rm RME}({\rm 2b}) = 4\, \pi \int_0^\infty dr\, r^2\, \rho^{\rm 2b}(r)   \, ,
\end{equation}
where $r$ is the interparticle distance, and ${\rm 2b}\,$=$\,$N2LO-$\Delta$, 
N2LO-OPE, and N2LO-CT.

Two-body transition densities, calculated for selected nuclei with variational
wave functions corresponding to the Hamiltonian models NV2+3-Ia (Ia) and NV2+3-Ia* (Ia*),
are presented in Fig.~\ref{fig:2panel}.  These models produce similar N2LO-$\Delta$ and
N3LO-OPE densities, since they are based on
the same underlying NV2 interaction and only differ in the NV3 interaction.  Specifically, what
differs is the strength $z_0$ of the contact current---linked to the
(contact) three-nucleon interaction via the relation in Eq.~(\ref{eq:ez0}).  Because of the
different methodologies adopted in constraining $c_D$ and $c_E$ (the LECs parameterizing
the contact piece of the three-nucleon interaction), $z_0$ turns out to be much larger for Ia*
than for Ia, see Table~\ref{tab:cdce}.  Since the contact current is proportional to $z_0$, this also explains why the
corresponding density for Ia* is much larger (in magnitude) than for Ia.  Note that they
are both negative.  As a consequence, the total density (black symbols in Fig.~\ref{fig:2panel})
develops a node at around 1 fm in the case of model Ia*.

Another interesting feature of Fig.~\ref{fig:2panel} is the difference between 
the N2LO-$\Delta$ and N3LO-OPE densities at separations $r\gtrsim 2$ fm
for the transitions in the larger systems, especially those involving the $A\,$=$\,8$ 
resonant states.   In the
limit of vanishing momentum transfer we are considering here, the corresponding
currents have the same operator structure~\cite{Baroni:2018fdn}, up to a momentum
dependent term, absent in the N2LO-$\Delta$ current, which, however, we have explicitly
verified to give a numerically small contribution by direct calculation.  Examination of Eqs.~(2.9)
and (2.10) of Ref.~\cite{Baroni:2018fdn} shows that this common operator structure involves
two independent correlations functions $I^{(1)}(\mu_{ij};\alpha_p)$ and $I^{(2)}(\mu_{ij};\alpha_p)$
with $\mu_{ij}\,$=$\,m_\pi\, r_{ij}$,
proportional to different combinations of LECs, denoted by $\alpha_1$ ($\alpha^\Delta_1$) 
and $\alpha_2$ ($\alpha^\Delta_2$) in the OPE ($\Delta$) current, with
\begin{eqnarray}
&&\frac{\alpha_1^\Delta}{\alpha_1} = \frac{c_4^\Delta}{c_4+1/(4\,m)} \approx 0.89 \ ,
\qquad \frac{\alpha_2^\Delta}{\alpha_2} = \frac{c_3^\Delta}{c_3} \approx 3.6 \ , \nonumber\\
&&\hspace{1cm}\frac{\alpha^\Delta_1}{\alpha_2^\Delta}= -\frac{1}{4} \ ,\qquad \qquad\qquad
\frac{\alpha_1}{\alpha_2} \approx -1.0
\end{eqnarray}
using the values in Tables~\ref{tab:lec1} and~\ref{tab:lec2} ($m$ is the average
nucleon mass).  Here, $c_3^\Delta\,$=$\,-h_A^2/(9 \, m_{\Delta N})$ and $c_4^\Delta\,$=$\,h_A^2/(18 \, m_{\Delta N})$.  Indeed, the N2LO-$\Delta$ current reads
\begin{equation}
{\bf j}^{\rm N2LO}_\Delta=
-\left({\bm \tau}_i \times {\bm \tau}_j\right)_a {\bm \sigma}_i \times {\bf O}^{(1)}_{ij}
- \tau_{j,a} \,{\bf O}^{(2)}_{ij}  +(i\rightleftharpoons j )\ ,
\end{equation}
with
\begin{equation}
{\bf O}^{(p)}_{ij}=I^{(1)}(\mu_{ij};\alpha_p^\Delta) \, {\bm \sigma}_j
+I^{(2)}(\mu_{ij};\alpha_p^\Delta) \,\hat{\bf r}_{ij}\,\,
 {\bm \sigma}_j\cdot \hat{\bf r}_{ij} \ .
\end{equation}
A similar expression holds for ${\bf j}^{\rm N3LO}_\pi$ with $\alpha_p$
replacing $\alpha^\Delta_p$.  There are
cancellations between the terms proportional to ${\bf O}^{(1)}_{ij}$
and ${\bf O}^{(2)}_{ij}$ in each of these currents, and these cancellations
are sensitive to the values of the ratios $\alpha_1^\Delta/\alpha_2^\Delta$
and  $\alpha_1/\alpha_2$, and to the overlap between the wave functions
of the states involved in the transition.

\begin{figure*}[bth]
\includegraphics[height=3.4in]{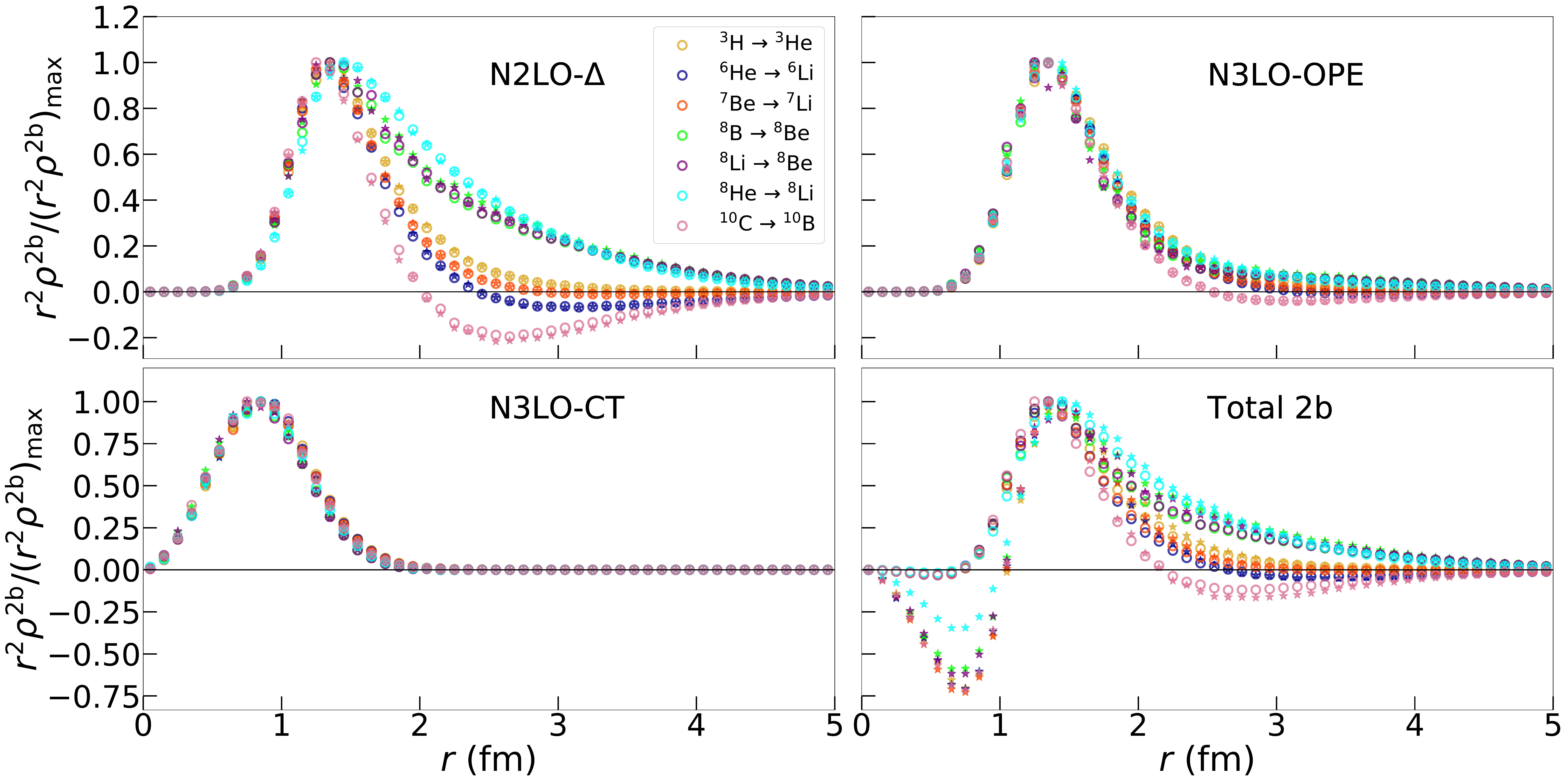}
\caption{(Color online)   Two-body transition density  for 
selected nuclei (see text for explanation). Different colors indicate different transitions. 
Results obtained from calculations using the NV2+3-Ia model are represented with open circles 
and results obtained from calculations employing the NV2+3-Ia* model results are represented by 
solid stars.}
\label{fig:2scaling}
\end{figure*}

To gain insight into how short-range physics impacts these weak transitions
across different nuclei,
we display in Fig.~\ref{fig:2scaling} the densities corresponding to the
individual two-body contributions, each normalized as
$4\pi r^2\rho^{\rm 2b}(r)/(4\pi r^2\rho^{\rm 2b})_{\rm max}$,
where $(4\pi r^2\rho^{\rm 2b})_{\rm max}$ is the maximum attained
value (in magnitude); so all curves peak at 1.  We also display the total
densities and note that, since both a positive peak and a negative 
valley are present in this case, each curve is normalized so that the
value of the positive peak is 1.

The universal behavior exhibited by the N2LO-$\Delta$, N3LO-OPE, N3LO-CT densities is
quite striking, as the curves corresponding to different nuclei and different Hamiltonian
models, essentially overlap for $r \lesssim 1/m_\pi$. (It is even
more striking when the weighing $r^2$ factor is not included).
Such behavior can be understood as follows.
\begin{figure*}[bth]
\includegraphics[width=3.7in]{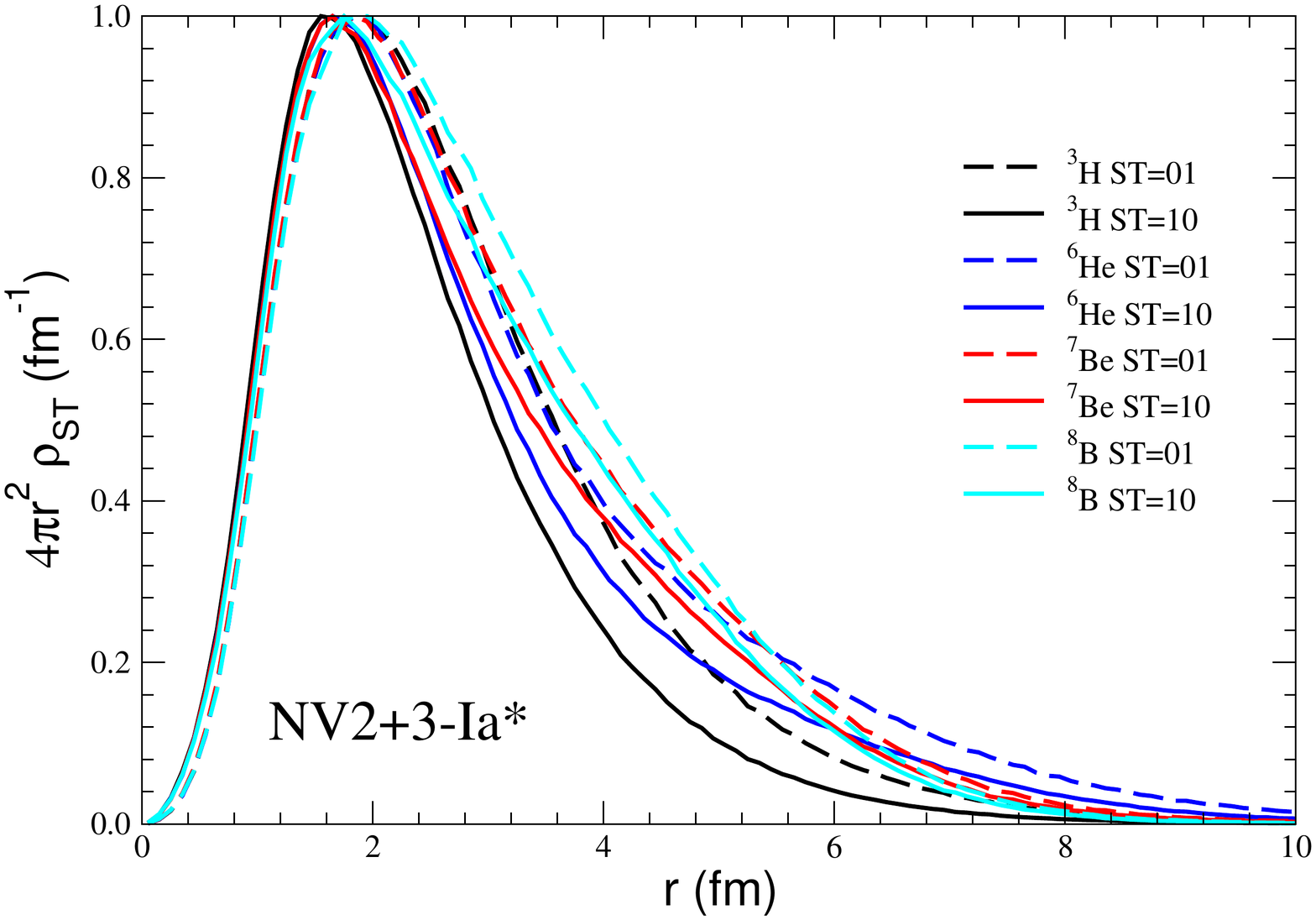}\includegraphics[width=3.7in]{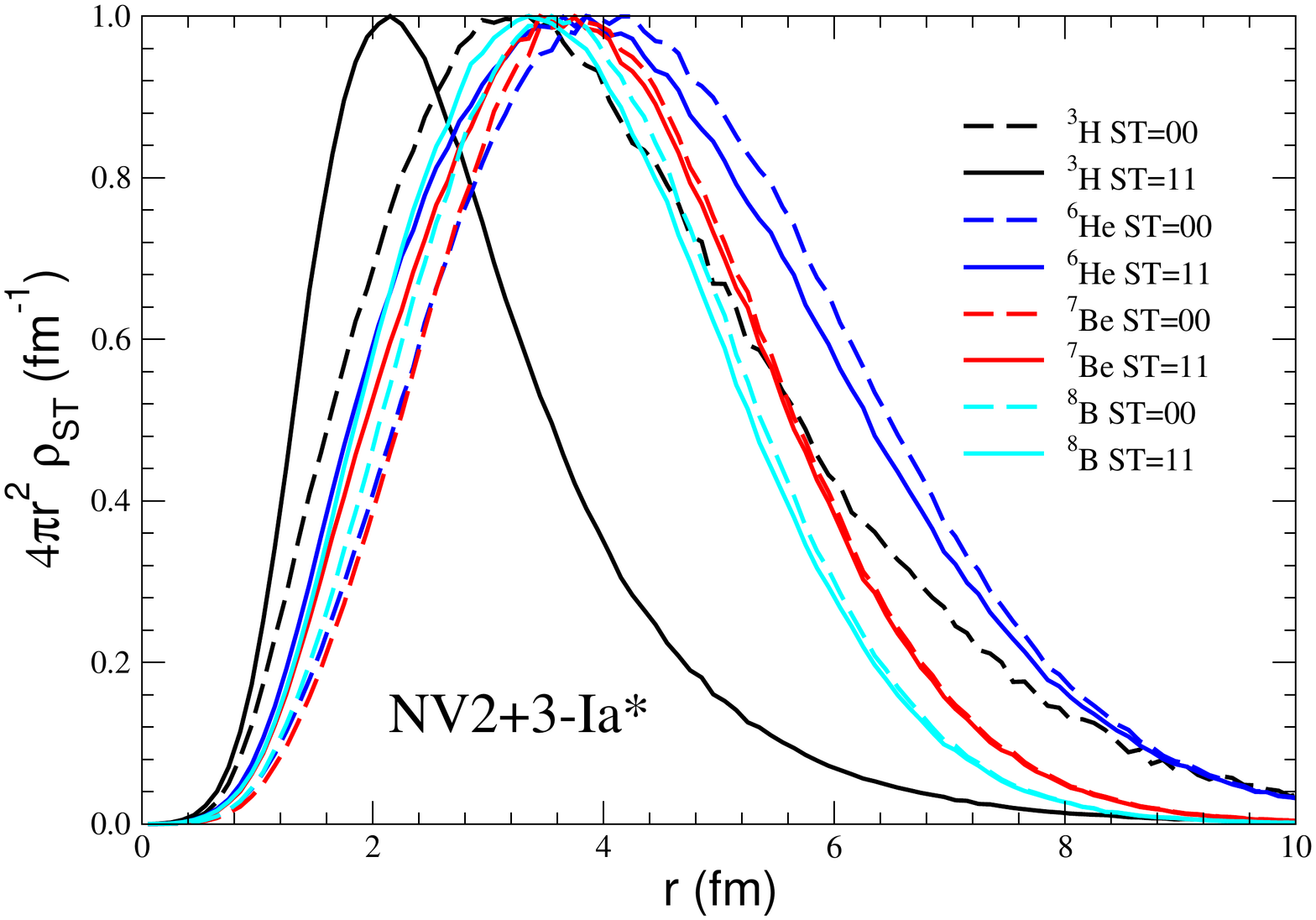}
\caption{(Color online) Pair densities in total spin/isospin $S/T\,$=$\,1/0$ and $0/1$ (left
panel) and $S/T\,$=$\,0/0$ and $1/1$ (right panel) obtained with VMC
wave functions corresponding to model Ia*.  All curves have been rescaled
so as peak at 1.  Only central values are shown; statistical Monte Carlo
errors are below a few \%. }
\label{fig:bob1}
\end{figure*}

In a charge-raising process the two-body weak transition operators
primarily convert a $pn$ pair with total spin/isospin $S/T\,$=$\,$1/0
($nn$ pair with $S/T\,$=$\,$0/1) to a $pp$ pair with $S/T\,$=$\,$0/1 ($pn$ pair with
$S/T\,$=$\,$1/0)~\cite{Schiavilla:1998je}---of course, similar considerations apply to
a charge lowering process.  These operators, at least in light systems, do
not couple $TT_z\,$=$\,10$ and $00$ to
$TT_z\,$=$\,11$ in a significant way, since P-wave
components are small in that case.  At  separations $\lesssim 1/m_\pi$, where
these transitions operators are most effective, the pair
wave functions with $S/T\,$=$\,$1/0 and 0/1
in different nuclei are similar in shape and only
differ by a scale factor~\cite{Forest:1996kp}---those corresponding to the Ia* model are illustrated in Fig.~\ref{fig:bob1}.  This is the origin for the universal
behavior observed in Fig.~\ref{fig:2scaling}.  Note that in Fig.~\ref{fig:bob1}
we also show the densities of pairs in $S/T$=0/0 and $1/1$, which do not scale.
A complete analysis and interpretation of these results---in particular, of the role played
by the tensor component of the nuclear interaction in shaping these
densities---can be found in Ref.~\cite{Forest:1996kp}.

However, at separations $r \gtrsim 1/m_\pi$, the N2LO-$\Delta$
and N3LO-OPE densities (especially the former) do not scale,
particularly in the case of the heavier systems with $A \geq 8$, presumably
due to delicate cancellations between the terms proportional to
the operators ${\bf O}^{(1)}_{ij}$ and ${\bf O}^{(2)}_{ij}$ present in
these currents (P-wave components in the wave functions of these systems may also
play a significant role).
As a matter of fact,
because of the different long-range behavior present in the $A\ge 8$ transitions (note, {\it e.g.},
how the N2LO-$\Delta$ density in the $^{10}$C transition assumes negative values at large 
separations, all the other densities being positive), and because of the rather large size of
the contact contribution in the Ia* model, the total densities exhibit nodes when calculated
with the starred interactions, which leads to non-trivial cancellations.

\section {Conclusions}
\label{sec:concl}

In this work, we reported on a detailed study of weak matrix elements
in $A\,$=$\,3$--10 systems based on chiral (two- and three-nucleon) interactions 
and associated (one- and two-body) axial currents at high orders in the chiral
expansion.  A summary of our results is displayed in Fig.~\ref{fig:rmes}. 
Agreement with the experimental data is obtained when correlated nuclear
wave functions are adopted.  For these transitions the contribution of corrections
beyond LO in the axial current is typically at the $\approx 2\%$ level of the value
calculated with the LO Gamow-Teller operator. These findings are in line with
those reported in the hybrid study of Ref.~\cite{Pastore:2017uwc} for the same
transitions.  Here, we also present calculations of matrix elements entering the
rates of the $^8$Li, $^8$B, and $^8$He beta decays.  These matrix elements are
found to be suppressed at LO, and N2LO and N3LO currents provide a large
correction ($\approx 20$--30\%) which is, however, insufficient to explain 
the experimental data. We attribute the large suppression at LO to the fact that
the Gamow-Teller operator is, in these  transitions, connecting large to small
components of the initial and final wave functions.  Improving on these
calculations will require the development of more sophisticated wave functions
with better constrained small components. 

Finally, we also reported on a careful analysis of one- and two-body transition
densities shown in Figs.~\ref{fig:1density7}--\ref{fig:2scaling}. The latter are especially
interesting because they allow us to understand  the spatial distributions of the
various two-body current operators, that is their behavior as functions of interparticle
distance.  We have shown that, for each set of interactions and consistent currents (either
NV2+3-Ia or NV2+3-Ia*), the two-body transition densities exhibit a universal behavior
at short distance across all nuclei we have considered in the present study. 

\section*{Acknowledgments}

We would like to thank V.\ Cirigliano, R. \ Charity,  W.\ Dekens, J. Engel, 
A.\ Hayes, E.\ Mereghetti, and L. \ Sobotka for useful discussions at various stages 
of this work. S.\ P. and G.\ K. would like to thank Grigor Sargsyan
for useful discussions and for sharing information on the experimental 
log$(ft)$ value of $^8$Li. The many-body calculations were performed
on the parallel computers of the Laboratory Computing Resource Center,
Argonne National Laboratory, the computers of the Argonne Leadership Computing
Facility (ALCF) via the 2019/2020 ALCC grant ``Low energy neutrino-nucleus interactions''
for the project NNInteractions.
The work of J.C., S.G., and~R.B.W.~has been supported by the NUclear Computational 
Low-Energy Initiative (NUCLEI) SciDAC project. This research is supported by the 
U.S.~Department of Energy, Office of Science,  Office of Nuclear Physics, under 
contracts DE-AC05-06OR23177 (R.S.) and DE-AC02-06CH11357 (R.B.W.), DE-AC52-06NA25396 (S.G. and J.C.)
and U.S.~Department of Energy funds through the FRIB Theory Alliance award DE-SC0013617 
(M.P. and S.P.).
\bibliography{beta}

\begin{thebibliography}{82}
\expandafter\ifx\csname natexlab\endcsname\relax\def\natexlab#1{#1}\fi
\expandafter\ifx\csname bibnamefont\endcsname\relax
  \def\bibnamefont#1{#1}\fi
\expandafter\ifx\csname bibfnamefont\endcsname\relax
  \def\bibfnamefont#1{#1}\fi
\expandafter\ifx\csname citenamefont\endcsname\relax
  \def\citenamefont#1{#1}\fi
\expandafter\ifx\csname url\endcsname\relax
  \def\url#1{\texttt{#1}}\fi
\expandafter\ifx\csname urlprefix\endcsname\relax\def\urlprefix{URL }\fi
\providecommand{\bibinfo}[2]{#2}
\providecommand{\eprint}[2][]{\url{#2}}

\bibitem[{\citenamefont{Piarulli et~al.}(2015)\citenamefont{Piarulli, Girlanda,
  Schiavilla, Navarro~Pérez, Amaro, and Ruiz~Arriola}}]{Piarulli:2014bda}
\bibinfo{author}{\bibfnamefont{M.}~\bibnamefont{Piarulli}},
  \bibinfo{author}{\bibfnamefont{L.}~\bibnamefont{Girlanda}},
  \bibinfo{author}{\bibfnamefont{R.}~\bibnamefont{Schiavilla}},
  \bibinfo{author}{\bibfnamefont{R.}~\bibnamefont{Navarro~Pérez}},
  \bibinfo{author}{\bibfnamefont{J.~E.} \bibnamefont{Amaro}}, \bibnamefont{and}
  \bibinfo{author}{\bibfnamefont{E.}~\bibnamefont{Ruiz~Arriola}},
  \bibinfo{journal}{Phys. Rev.} \textbf{\bibinfo{volume}{C91}},
  \bibinfo{pages}{024003} (\bibinfo{year}{2015}), \eprint{1412.6446}.

\bibitem[{\citenamefont{Piarulli et~al.}(2016)\citenamefont{Piarulli, Girlanda,
  Schiavilla, Kievsky, Lovato, Marcucci, Pieper, Viviani, and
  Wiringa}}]{Piarulli:2016vel}
\bibinfo{author}{\bibfnamefont{M.}~\bibnamefont{Piarulli}},
  \bibinfo{author}{\bibfnamefont{L.}~\bibnamefont{Girlanda}},
  \bibinfo{author}{\bibfnamefont{R.}~\bibnamefont{Schiavilla}},
  \bibinfo{author}{\bibfnamefont{A.}~\bibnamefont{Kievsky}},
  \bibinfo{author}{\bibfnamefont{A.}~\bibnamefont{Lovato}},
  \bibinfo{author}{\bibfnamefont{L.~E.} \bibnamefont{Marcucci}},
  \bibinfo{author}{\bibfnamefont{S.~C.} \bibnamefont{Pieper}},
  \bibinfo{author}{\bibfnamefont{M.}~\bibnamefont{Viviani}}, \bibnamefont{and}
  \bibinfo{author}{\bibfnamefont{R.~B.} \bibnamefont{Wiringa}},
  \bibinfo{journal}{Phys. Rev.} \textbf{\bibinfo{volume}{C94}},
  \bibinfo{pages}{054007} (\bibinfo{year}{2016}), \eprint{1606.06335}.

\bibitem[{\citenamefont{Baroni et~al.}(2016{\natexlab{a}})\citenamefont{Baroni,
  Girlanda, Kievsky, Marcucci, Schiavilla, and Viviani}}]{Baroni:2016xll}
\bibinfo{author}{\bibfnamefont{A.}~\bibnamefont{Baroni}},
  \bibinfo{author}{\bibfnamefont{L.}~\bibnamefont{Girlanda}},
  \bibinfo{author}{\bibfnamefont{A.}~\bibnamefont{Kievsky}},
  \bibinfo{author}{\bibfnamefont{L.~E.} \bibnamefont{Marcucci}},
  \bibinfo{author}{\bibfnamefont{R.}~\bibnamefont{Schiavilla}},
  \bibnamefont{and} \bibinfo{author}{\bibfnamefont{M.}~\bibnamefont{Viviani}},
  \bibinfo{journal}{Phys. Rev.} \textbf{\bibinfo{volume}{C94}},
  \bibinfo{pages}{024003} (\bibinfo{year}{2016}{\natexlab{a}}),
  \bibinfo{note}{[Erratum: Phys. Rev.C95,no.5,059902(2017)]},
  \eprint{1605.01620}.

\bibitem[{\citenamefont{Baroni et~al.}(2018)}]{Baroni:2018fdn}
\bibinfo{author}{\bibfnamefont{A.}~\bibnamefont{Baroni}} \bibnamefont{et~al.},
  \bibinfo{journal}{Phys. Rev.} \textbf{\bibinfo{volume}{C98}},
  \bibinfo{pages}{044003} (\bibinfo{year}{2018}), \eprint{1806.10245}.

\bibitem[{\citenamefont{Baroni et~al.}(2016{\natexlab{b}})\citenamefont{Baroni,
  Girlanda, Pastore, Schiavilla, and Viviani}}]{Baroni:2015uza}
\bibinfo{author}{\bibfnamefont{A.}~\bibnamefont{Baroni}},
  \bibinfo{author}{\bibfnamefont{L.}~\bibnamefont{Girlanda}},
  \bibinfo{author}{\bibfnamefont{S.}~\bibnamefont{Pastore}},
  \bibinfo{author}{\bibfnamefont{R.}~\bibnamefont{Schiavilla}},
  \bibnamefont{and} \bibinfo{author}{\bibfnamefont{M.}~\bibnamefont{Viviani}},
  \bibinfo{journal}{Phys. Rev.} \textbf{\bibinfo{volume}{C93}},
  \bibinfo{pages}{015501} (\bibinfo{year}{2016}{\natexlab{b}}),
  \bibinfo{note}{[Erratum: Phys. Rev.C95,no.5,059901(2017)]},
  \eprint{1509.07039}.

\bibitem[{\citenamefont{Pastore
  et~al.}(2018{\natexlab{a}})\citenamefont{Pastore, Baroni, Carlson, Gandolfi,
  Pieper, Schiavilla, and Wiringa}}]{Pastore:2017uwc}
\bibinfo{author}{\bibfnamefont{S.}~\bibnamefont{Pastore}},
  \bibinfo{author}{\bibfnamefont{A.}~\bibnamefont{Baroni}},
  \bibinfo{author}{\bibfnamefont{J.}~\bibnamefont{Carlson}},
  \bibinfo{author}{\bibfnamefont{S.}~\bibnamefont{Gandolfi}},
  \bibinfo{author}{\bibfnamefont{S.~C.} \bibnamefont{Pieper}},
  \bibinfo{author}{\bibfnamefont{R.}~\bibnamefont{Schiavilla}},
  \bibnamefont{and} \bibinfo{author}{\bibfnamefont{R.~B.}
  \bibnamefont{Wiringa}}, \bibinfo{journal}{Phys. Rev.}
  \textbf{\bibinfo{volume}{C97}}, \bibinfo{pages}{022501}
  (\bibinfo{year}{2018}{\natexlab{a}}), \eprint{1709.03592}.

\bibitem[{\citenamefont{Wiringa et~al.}(1995)\citenamefont{Wiringa, Stoks, and
  Schiavilla}}]{Wiringa:1994wb}
\bibinfo{author}{\bibfnamefont{R.~B.} \bibnamefont{Wiringa}},
  \bibinfo{author}{\bibfnamefont{V.~G.~J.} \bibnamefont{Stoks}},
  \bibnamefont{and}
  \bibinfo{author}{\bibfnamefont{R.}~\bibnamefont{Schiavilla}},
  \bibinfo{journal}{Phys. Rev.} \textbf{\bibinfo{volume}{C51}},
  \bibinfo{pages}{38} (\bibinfo{year}{1995}), \eprint{nucl-th/9408016}.

\bibitem[{\citenamefont{Pieper}(2008)}]{doi:10.1063/1.2932280}
\bibinfo{author}{\bibfnamefont{S.~C.} \bibnamefont{Pieper}},
  \bibinfo{journal}{AIP Conference Proceedings}
  \textbf{\bibinfo{volume}{1011}}, \bibinfo{pages}{143} (\bibinfo{year}{2008}),
  \eprint{https://aip.scitation.org/doi/pdf/10.1063/1.2932280},
  \urlprefix\url{https://aip.scitation.org/doi/abs/10.1063/1.2932280}.

\bibitem[{\citenamefont{Gysbers et~al.}(2019)}]{Gysbers:2019uyb}
\bibinfo{author}{\bibfnamefont{P.}~\bibnamefont{Gysbers}} \bibnamefont{et~al.},
  \bibinfo{journal}{Nature Phys.} \textbf{\bibinfo{volume}{15}},
  \bibinfo{pages}{428} (\bibinfo{year}{2019}), \eprint{1903.00047}.

\bibitem[{\citenamefont{Gazit et~al.}(2009{\natexlab{a}})\citenamefont{Gazit,
  Vaintraub, and Barnea}}]{Gazit:2009kz}
\bibinfo{author}{\bibfnamefont{D.}~\bibnamefont{Gazit}},
  \bibinfo{author}{\bibfnamefont{S.}~\bibnamefont{Vaintraub}},
  \bibnamefont{and} \bibinfo{author}{\bibfnamefont{N.}~\bibnamefont{Barnea}},
  in \emph{\bibinfo{booktitle}{{Particles and nuclei. Proceedings, 18th
  International Conference, PANIC08, Eilat, Israel, November 9-14, 2008}}}
  (\bibinfo{year}{2009}{\natexlab{a}}), pp. \bibinfo{pages}{1005--1007},
  \eprint{0901.2670}.

\bibitem[{\citenamefont{Gardestig and Phillips}(2006)}]{Gardestig:2006hj}
\bibinfo{author}{\bibfnamefont{A.}~\bibnamefont{Gardestig}} \bibnamefont{and}
  \bibinfo{author}{\bibfnamefont{D.~R.} \bibnamefont{Phillips}},
  \bibinfo{journal}{Phys. Rev. Lett.} \textbf{\bibinfo{volume}{96}},
  \bibinfo{pages}{232301} (\bibinfo{year}{2006}), \eprint{nucl-th/0603045}.

\bibitem[{\citenamefont{Gazit et~al.}(2009{\natexlab{b}})\citenamefont{Gazit,
  Quaglioni, and Navratil}}]{Gazit:2008ma}
\bibinfo{author}{\bibfnamefont{D.}~\bibnamefont{Gazit}},
  \bibinfo{author}{\bibfnamefont{S.}~\bibnamefont{Quaglioni}},
  \bibnamefont{and} \bibinfo{author}{\bibfnamefont{P.}~\bibnamefont{Navratil}},
  \bibinfo{journal}{Phys. Rev. Lett.} \textbf{\bibinfo{volume}{103}},
  \bibinfo{pages}{102502} (\bibinfo{year}{2009}{\natexlab{b}}),
  \bibinfo{note}{[Erratum: Phys. Rev. Lett. 122, 029901 (2019)]}.

\bibitem[{\citenamefont{Schiavilla}(2017)}]{Schiavilla:2017}
\bibinfo{author}{\bibfnamefont{R.}~\bibnamefont{Schiavilla}},
  \bibinfo{journal}{unpublished}  (\bibinfo{year}{2017}).

\bibitem[{\citenamefont{Gonzalez-Alonso
  et~al.}(2019)\citenamefont{Gonzalez-Alonso, Naviliat-Cuncic, and
  Severijns}}]{Gonzalez-Alonso:2018omy}
\bibinfo{author}{\bibfnamefont{M.}~\bibnamefont{Gonzalez-Alonso}},
  \bibinfo{author}{\bibfnamefont{O.}~\bibnamefont{Naviliat-Cuncic}},
  \bibnamefont{and}
  \bibinfo{author}{\bibfnamefont{N.}~\bibnamefont{Severijns}},
  \bibinfo{journal}{Prog. Part. Nucl. Phys.} \textbf{\bibinfo{volume}{104}},
  \bibinfo{pages}{165} (\bibinfo{year}{2019}), \eprint{1803.08732}.

\bibitem[{\citenamefont{Engel and Menendez}(2017)}]{Engel:2016xgb}
\bibinfo{author}{\bibfnamefont{J.}~\bibnamefont{Engel}} \bibnamefont{and}
  \bibinfo{author}{\bibfnamefont{J.}~\bibnamefont{Menendez}},
  \bibinfo{journal}{Rept. Prog. Phys.} \textbf{\bibinfo{volume}{80}},
  \bibinfo{pages}{046301} (\bibinfo{year}{2017}), \eprint{1610.06548}.

\bibitem[{\citenamefont{Carlson et~al.}(2015)\citenamefont{Carlson, Gandolfi,
  Pederiva, Pieper, Schiavilla, Schmidt, and Wiringa}}]{Carlson:2014vla}
\bibinfo{author}{\bibfnamefont{J.}~\bibnamefont{Carlson}},
  \bibinfo{author}{\bibfnamefont{S.}~\bibnamefont{Gandolfi}},
  \bibinfo{author}{\bibfnamefont{F.}~\bibnamefont{Pederiva}},
  \bibinfo{author}{\bibfnamefont{S.~C.} \bibnamefont{Pieper}},
  \bibinfo{author}{\bibfnamefont{R.}~\bibnamefont{Schiavilla}},
  \bibinfo{author}{\bibfnamefont{K.~E.} \bibnamefont{Schmidt}},
  \bibnamefont{and} \bibinfo{author}{\bibfnamefont{R.~B.}
  \bibnamefont{Wiringa}}, \bibinfo{journal}{Rev. Mod. Phys.}
  \textbf{\bibinfo{volume}{87}}, \bibinfo{pages}{1067} (\bibinfo{year}{2015}),
  \eprint{1412.3081}.

\bibitem[{\citenamefont{Lynn et~al.}(2019)\citenamefont{Lynn, Tews, Gandolfi,
  and Lovato}}]{Lynn:2019rdt}
\bibinfo{author}{\bibfnamefont{J.~E.} \bibnamefont{Lynn}},
  \bibinfo{author}{\bibfnamefont{I.}~\bibnamefont{Tews}},
  \bibinfo{author}{\bibfnamefont{S.}~\bibnamefont{Gandolfi}}, \bibnamefont{and}
  \bibinfo{author}{\bibfnamefont{A.}~\bibnamefont{Lovato}},
  \bibinfo{journal}{Submitted to: Ann. Rev. Nucl. Part. Sci.}
  (\bibinfo{year}{2019}), \eprint{1901.04868}.

\bibitem[{\citenamefont{Wiringa}(1991)}]{Wiringa:1991kp}
\bibinfo{author}{\bibfnamefont{R.~B.} \bibnamefont{Wiringa}},
  \bibinfo{journal}{Phys. Rev.} \textbf{\bibinfo{volume}{C43}},
  \bibinfo{pages}{1585} (\bibinfo{year}{1991}).

\bibitem[{\citenamefont{Pudliner et~al.}(1997)\citenamefont{Pudliner,
  Pandharipande, Carlson, Pieper, and Wiringa}}]{Pudliner:1997ck}
\bibinfo{author}{\bibfnamefont{B.~S.} \bibnamefont{Pudliner}},
  \bibinfo{author}{\bibfnamefont{V.~R.} \bibnamefont{Pandharipande}},
  \bibinfo{author}{\bibfnamefont{J.}~\bibnamefont{Carlson}},
  \bibinfo{author}{\bibfnamefont{S.~C.} \bibnamefont{Pieper}},
  \bibnamefont{and} \bibinfo{author}{\bibfnamefont{R.~B.}
  \bibnamefont{Wiringa}}, \bibinfo{journal}{Phys. Rev.}
  \textbf{\bibinfo{volume}{C56}}, \bibinfo{pages}{1720} (\bibinfo{year}{1997}),
  \eprint{nucl-th/9705009}.

\bibitem[{\citenamefont{Nollett et~al.}(2001)\citenamefont{Nollett, Wiringa,
  and Schiavilla}}]{Nollett:2000ch}
\bibinfo{author}{\bibfnamefont{K.~M.} \bibnamefont{Nollett}},
  \bibinfo{author}{\bibfnamefont{R.~B.} \bibnamefont{Wiringa}},
  \bibnamefont{and}
  \bibinfo{author}{\bibfnamefont{R.}~\bibnamefont{Schiavilla}},
  \bibinfo{journal}{Phys. Rev.} \textbf{\bibinfo{volume}{C63}},
  \bibinfo{pages}{024003} (\bibinfo{year}{2001}), \eprint{nucl-th/0006064}.

\bibitem[{\citenamefont{Nollett}(2001)}]{Nollett:2001ub}
\bibinfo{author}{\bibfnamefont{K.~M.} \bibnamefont{Nollett}},
  \bibinfo{journal}{Phys. Rev.} \textbf{\bibinfo{volume}{C63}},
  \bibinfo{pages}{054002} (\bibinfo{year}{2001}), \eprint{nucl-th/0102022}.

\bibitem[{\citenamefont{Metropolis et~al.}(1953)\citenamefont{Metropolis,
  Rosenbluth, Rosenbluth, Teller, and Teller}}]{Metropolis:1953am}
\bibinfo{author}{\bibfnamefont{N.}~\bibnamefont{Metropolis}},
  \bibinfo{author}{\bibfnamefont{A.~W.} \bibnamefont{Rosenbluth}},
  \bibinfo{author}{\bibfnamefont{M.~N.} \bibnamefont{Rosenbluth}},
  \bibinfo{author}{\bibfnamefont{A.~H.} \bibnamefont{Teller}},
  \bibnamefont{and} \bibinfo{author}{\bibfnamefont{E.}~\bibnamefont{Teller}},
  \bibinfo{journal}{J. Chem. Phys.} \textbf{\bibinfo{volume}{21}},
  \bibinfo{pages}{1087} (\bibinfo{year}{1953}).

\bibitem[{\citenamefont{Pervin et~al.}(2007)\citenamefont{Pervin, Pieper, and
  Wiringa}}]{Pervin:2007sc}
\bibinfo{author}{\bibfnamefont{M.}~\bibnamefont{Pervin}},
  \bibinfo{author}{\bibfnamefont{S.~C.} \bibnamefont{Pieper}},
  \bibnamefont{and} \bibinfo{author}{\bibfnamefont{R.~B.}
  \bibnamefont{Wiringa}}, \bibinfo{journal}{Phys. Rev.}
  \textbf{\bibinfo{volume}{C76}}, \bibinfo{pages}{064319}
  (\bibinfo{year}{2007}), \eprint{0710.1265}.

\bibitem[{\citenamefont{Krebs et~al.}(2007)\citenamefont{Krebs, Epelbaum, and
  Meissner}}]{Krebs:2007rh}
\bibinfo{author}{\bibfnamefont{H.}~\bibnamefont{Krebs}},
  \bibinfo{author}{\bibfnamefont{E.}~\bibnamefont{Epelbaum}}, \bibnamefont{and}
  \bibinfo{author}{\bibfnamefont{U.-G.} \bibnamefont{Meissner}},
  \bibinfo{journal}{Eur. Phys. J.} \textbf{\bibinfo{volume}{A32}},
  \bibinfo{pages}{127} (\bibinfo{year}{2007}), \eprint{nucl-th/0703087}.

\bibitem[{\citenamefont{Navarro~Pérez
  et~al.}(2013)\citenamefont{Navarro~Pérez, Amaro, and
  Ruiz~Arriola}}]{Perez:2013jpa}
\bibinfo{author}{\bibfnamefont{R.}~\bibnamefont{Navarro~Pérez}},
  \bibinfo{author}{\bibfnamefont{J.~E.} \bibnamefont{Amaro}}, \bibnamefont{and}
  \bibinfo{author}{\bibfnamefont{E.}~\bibnamefont{Ruiz~Arriola}},
  \bibinfo{journal}{Phys. Rev.} \textbf{\bibinfo{volume}{C88}},
  \bibinfo{pages}{064002} (\bibinfo{year}{2013}), \bibinfo{note}{[Erratum:
  Phys. Rev.C91,no.2,029901(2015)]}, \eprint{1310.2536}.

\bibitem[{\citenamefont{Navarro~Pérez
  et~al.}(2014)\citenamefont{Navarro~Pérez, Amaro, and
  Ruiz~Arriola}}]{Perez:2013oba}
\bibinfo{author}{\bibfnamefont{R.}~\bibnamefont{Navarro~Pérez}},
  \bibinfo{author}{\bibfnamefont{J.~E.} \bibnamefont{Amaro}}, \bibnamefont{and}
  \bibinfo{author}{\bibfnamefont{E.}~\bibnamefont{Ruiz~Arriola}},
  \bibinfo{journal}{Phys. Rev. C} \textbf{\bibinfo{volume}{89}},
  \bibinfo{pages}{024004} (\bibinfo{year}{2014}), \eprint{1310.6972}.

\bibitem[{\citenamefont{Navarro~Perez et~al.}(2014)\citenamefont{Navarro~Perez,
  Amaro, and Ruiz~Arriola}}]{Perez:2014yla}
\bibinfo{author}{\bibfnamefont{R.}~\bibnamefont{Navarro~Perez}},
  \bibinfo{author}{\bibfnamefont{J.~E.} \bibnamefont{Amaro}}, \bibnamefont{and}
  \bibinfo{author}{\bibfnamefont{E.}~\bibnamefont{Ruiz~Arriola}},
  \bibinfo{journal}{Phys. Rev. C} \textbf{\bibinfo{volume}{89}},
  \bibinfo{pages}{064006} (\bibinfo{year}{2014}), \eprint{1404.0314}.

\bibitem[{\citenamefont{Green}(1976)}]{Green:1976wx}
\bibinfo{author}{\bibfnamefont{A.~M.} \bibnamefont{Green}},
  \bibinfo{journal}{Rept. Prog. Phys.} \textbf{\bibinfo{volume}{39}},
  \bibinfo{pages}{1109} (\bibinfo{year}{1976}).

\bibitem[{\citenamefont{Arndt et~al.}(1994)\citenamefont{Arndt, Workman, and
  Pavan}}]{Arndt:1994bu}
\bibinfo{author}{\bibfnamefont{R.~A.} \bibnamefont{Arndt}},
  \bibinfo{author}{\bibfnamefont{R.~L.} \bibnamefont{Workman}},
  \bibnamefont{and} \bibinfo{author}{\bibfnamefont{M.~M.} \bibnamefont{Pavan}},
  \bibinfo{journal}{Phys. Rev.} \textbf{\bibinfo{volume}{C49}},
  \bibinfo{pages}{2729} (\bibinfo{year}{1994}).

\bibitem[{\citenamefont{Stoks et~al.}(1993)\citenamefont{Stoks, Timmermans, and
  de~Swart}}]{Stoks:1992ja}
\bibinfo{author}{\bibfnamefont{V.~G.~J.} \bibnamefont{Stoks}},
  \bibinfo{author}{\bibfnamefont{R.}~\bibnamefont{Timmermans}},
  \bibnamefont{and} \bibinfo{author}{\bibfnamefont{J.~J.}
  \bibnamefont{de~Swart}}, \bibinfo{journal}{Phys. Rev.}
  \textbf{\bibinfo{volume}{C47}}, \bibinfo{pages}{512} (\bibinfo{year}{1993}),
  \eprint{nucl-th/9211007}.

\bibitem[{\citenamefont{Tanabashi et~al.}(2018)\citenamefont{Tanabashi,
  Hagiwara, Hikasa, Nakamura, Sumino, Takahashi, Tanaka, Agashe, Aielli, Amsler
  et~al.}}]{PDG}
\bibinfo{author}{\bibfnamefont{M.}~\bibnamefont{Tanabashi}},
  \bibinfo{author}{\bibfnamefont{K.}~\bibnamefont{Hagiwara}},
  \bibinfo{author}{\bibfnamefont{K.}~\bibnamefont{Hikasa}},
  \bibinfo{author}{\bibfnamefont{K.}~\bibnamefont{Nakamura}},
  \bibinfo{author}{\bibfnamefont{Y.}~\bibnamefont{Sumino}},
  \bibinfo{author}{\bibfnamefont{F.}~\bibnamefont{Takahashi}},
  \bibinfo{author}{\bibfnamefont{J.}~\bibnamefont{Tanaka}},
  \bibinfo{author}{\bibfnamefont{K.}~\bibnamefont{Agashe}},
  \bibinfo{author}{\bibfnamefont{G.}~\bibnamefont{Aielli}},
  \bibinfo{author}{\bibfnamefont{C.}~\bibnamefont{Amsler}},
  \bibnamefont{et~al.} (\bibinfo{collaboration}{Particle Data Group}),
  \bibinfo{journal}{Phys. Rev. D} \textbf{\bibinfo{volume}{98}},
  \bibinfo{pages}{030001} (\bibinfo{year}{2018}),
  \urlprefix\url{https://link.aps.org/doi/10.1103/PhysRevD.98.030001}.

\bibitem[{\citenamefont{van Kolck}(1994)}]{vanKolck:1994yi}
\bibinfo{author}{\bibfnamefont{U.}~\bibnamefont{van Kolck}},
  \bibinfo{journal}{Phys. Rev.} \textbf{\bibinfo{volume}{C49}},
  \bibinfo{pages}{2932} (\bibinfo{year}{1994}).

\bibitem[{\citenamefont{Epelbaum et~al.}(2002)\citenamefont{Epelbaum, Nogga,
  Gloeckle, Kamada, Meissner, and Witala}}]{Epelbaum:2002vt}
\bibinfo{author}{\bibfnamefont{E.}~\bibnamefont{Epelbaum}},
  \bibinfo{author}{\bibfnamefont{A.}~\bibnamefont{Nogga}},
  \bibinfo{author}{\bibfnamefont{W.}~\bibnamefont{Gloeckle}},
  \bibinfo{author}{\bibfnamefont{H.}~\bibnamefont{Kamada}},
  \bibinfo{author}{\bibfnamefont{U.~G.} \bibnamefont{Meissner}},
  \bibnamefont{and} \bibinfo{author}{\bibfnamefont{H.}~\bibnamefont{Witala}},
  \bibinfo{journal}{Phys. Rev.} \textbf{\bibinfo{volume}{C66}},
  \bibinfo{pages}{064001} (\bibinfo{year}{2002}), \eprint{nucl-th/0208023}.

\bibitem[{\citenamefont{Lynn et~al.}(2016)\citenamefont{Lynn, Tews, Carlson,
  Gandolfi, Gezerlis, Schmidt, and Schwenk}}]{Lynn:2015jua}
\bibinfo{author}{\bibfnamefont{J.~E.} \bibnamefont{Lynn}},
  \bibinfo{author}{\bibfnamefont{I.}~\bibnamefont{Tews}},
  \bibinfo{author}{\bibfnamefont{J.}~\bibnamefont{Carlson}},
  \bibinfo{author}{\bibfnamefont{S.}~\bibnamefont{Gandolfi}},
  \bibinfo{author}{\bibfnamefont{A.}~\bibnamefont{Gezerlis}},
  \bibinfo{author}{\bibfnamefont{K.~E.} \bibnamefont{Schmidt}},
  \bibnamefont{and} \bibinfo{author}{\bibfnamefont{A.}~\bibnamefont{Schwenk}},
  \bibinfo{journal}{Phys. Rev. Lett.} \textbf{\bibinfo{volume}{116}},
  \bibinfo{pages}{062501} (\bibinfo{year}{2016}), \eprint{1509.03470}.

\bibitem[{\citenamefont{Tews et~al.}(2016)\citenamefont{Tews, Gandolfi,
  Gezerlis, and Schwenk}}]{Tews:2015ufa}
\bibinfo{author}{\bibfnamefont{I.}~\bibnamefont{Tews}},
  \bibinfo{author}{\bibfnamefont{S.}~\bibnamefont{Gandolfi}},
  \bibinfo{author}{\bibfnamefont{A.}~\bibnamefont{Gezerlis}}, \bibnamefont{and}
  \bibinfo{author}{\bibfnamefont{A.}~\bibnamefont{Schwenk}},
  \bibinfo{journal}{Phys. Rev.} \textbf{\bibinfo{volume}{C93}},
  \bibinfo{pages}{024305} (\bibinfo{year}{2016}), \eprint{1507.05561}.

\bibitem[{\citenamefont{Lynn et~al.}(2017)\citenamefont{Lynn, Tews, Carlson,
  Gandolfi, Gezerlis, Schmidt, and Schwenk}}]{Lynn:2017fxg}
\bibinfo{author}{\bibfnamefont{J.~E.} \bibnamefont{Lynn}},
  \bibinfo{author}{\bibfnamefont{I.}~\bibnamefont{Tews}},
  \bibinfo{author}{\bibfnamefont{J.}~\bibnamefont{Carlson}},
  \bibinfo{author}{\bibfnamefont{S.}~\bibnamefont{Gandolfi}},
  \bibinfo{author}{\bibfnamefont{A.}~\bibnamefont{Gezerlis}},
  \bibinfo{author}{\bibfnamefont{K.~E.} \bibnamefont{Schmidt}},
  \bibnamefont{and} \bibinfo{author}{\bibfnamefont{A.}~\bibnamefont{Schwenk}},
  \bibinfo{journal}{Phys. Rev.} \textbf{\bibinfo{volume}{C96}},
  \bibinfo{pages}{054007} (\bibinfo{year}{2017}), \eprint{1706.07668}.

\bibitem[{\citenamefont{Piarulli et~al.}(2018)}]{Piarulli:2017dwd}
\bibinfo{author}{\bibfnamefont{M.}~\bibnamefont{Piarulli}}
  \bibnamefont{et~al.}, \bibinfo{journal}{Phys. Rev. Lett.}
  \textbf{\bibinfo{volume}{120}}, \bibinfo{pages}{052503}
  (\bibinfo{year}{2018}), \eprint{1707.02883}.

\bibitem[{\citenamefont{Marcucci et~al.}(2012)\citenamefont{Marcucci, Kievsky,
  Rosati, Schiavilla, and Viviani}}]{Marcucci:2011jm}
\bibinfo{author}{\bibfnamefont{L.~E.} \bibnamefont{Marcucci}},
  \bibinfo{author}{\bibfnamefont{A.}~\bibnamefont{Kievsky}},
  \bibinfo{author}{\bibfnamefont{S.}~\bibnamefont{Rosati}},
  \bibinfo{author}{\bibfnamefont{R.}~\bibnamefont{Schiavilla}},
  \bibnamefont{and} \bibinfo{author}{\bibfnamefont{M.}~\bibnamefont{Viviani}},
  \bibinfo{journal}{Phys. Rev. Lett.} \textbf{\bibinfo{volume}{108}},
  \bibinfo{pages}{052502} (\bibinfo{year}{2012}), \bibinfo{note}{[Erratum:
  Phys. Rev. Lett.121,no.4,049901(2018)]}, \eprint{1109.5563}.

\bibitem[{\citenamefont{Cirigliano et~al.}(2019)\citenamefont{Cirigliano,
  Dekens, De~Vries, Graesser, Mereghetti, Pastore, Piarulli, Van~Kolck, and
  Wiringa}}]{Cirigliano:2019vdj}
\bibinfo{author}{\bibfnamefont{V.}~\bibnamefont{Cirigliano}},
  \bibinfo{author}{\bibfnamefont{W.}~\bibnamefont{Dekens}},
  \bibinfo{author}{\bibfnamefont{J.}~\bibnamefont{De~Vries}},
  \bibinfo{author}{\bibfnamefont{M.~L.} \bibnamefont{Graesser}},
  \bibinfo{author}{\bibfnamefont{E.}~\bibnamefont{Mereghetti}},
  \bibinfo{author}{\bibfnamefont{S.}~\bibnamefont{Pastore}},
  \bibinfo{author}{\bibfnamefont{M.}~\bibnamefont{Piarulli}},
  \bibinfo{author}{\bibfnamefont{U.}~\bibnamefont{Van~Kolck}},
  \bibnamefont{and} \bibinfo{author}{\bibfnamefont{R.~B.}
  \bibnamefont{Wiringa}} (\bibinfo{year}{2019}), \eprint{1907.11254}.

\bibitem[{\citenamefont{Piarulli and Tews}(2020)}]{10.3389/fphy.2019.00245}
\bibinfo{author}{\bibfnamefont{M.}~\bibnamefont{Piarulli}} \bibnamefont{and}
  \bibinfo{author}{\bibfnamefont{I.}~\bibnamefont{Tews}},
  \bibinfo{journal}{Frontiers in Physics} \textbf{\bibinfo{volume}{7}},
  \bibinfo{pages}{245} (\bibinfo{year}{2020}), ISSN \bibinfo{issn}{2296-424X},
  \urlprefix\url{https://www.frontiersin.org/article/10.3389/fphy.2019.00245}.

\bibitem[{\citenamefont{Gandolfi et~al.}(2020)\citenamefont{Gandolfi,
  Lonardoni, Lovato, and Piarulli}}]{Gandolfi:2020pbj}
\bibinfo{author}{\bibfnamefont{S.}~\bibnamefont{Gandolfi}},
  \bibinfo{author}{\bibfnamefont{D.}~\bibnamefont{Lonardoni}},
  \bibinfo{author}{\bibfnamefont{A.}~\bibnamefont{Lovato}}, \bibnamefont{and}
  \bibinfo{author}{\bibfnamefont{M.}~\bibnamefont{Piarulli}}
  (\bibinfo{year}{2020}), \eprint{2001.01374}.

\bibitem[{\citenamefont{Schmidt and Fantoni}(1999)}]{Schmidt:1999lik}
\bibinfo{author}{\bibfnamefont{K.~E.} \bibnamefont{Schmidt}} \bibnamefont{and}
  \bibinfo{author}{\bibfnamefont{S.}~\bibnamefont{Fantoni}},
  \bibinfo{journal}{Phys. Lett.} \textbf{\bibinfo{volume}{B446}},
  \bibinfo{pages}{99} (\bibinfo{year}{1999}).

\bibitem[{\citenamefont{Day}(1967)}]{bbg1}
\bibinfo{author}{\bibfnamefont{B.~D.} \bibnamefont{Day}},
  \bibinfo{journal}{Rev. Mod. Phys.} \textbf{\bibinfo{volume}{39}},
  \bibinfo{pages}{719} (\bibinfo{year}{1967}).

\bibitem[{\citenamefont{Baldo and Burgio}(2012)}]{bbg2}
\bibinfo{author}{\bibfnamefont{M.}~\bibnamefont{Baldo}} \bibnamefont{and}
  \bibinfo{author}{\bibfnamefont{G.~F.} \bibnamefont{Burgio}},
  \bibinfo{journal}{Rep. Progr. Phys.} \textbf{\bibinfo{volume}{75}},
  \bibinfo{pages}{026301} (\bibinfo{year}{2012}).

\bibitem[{\citenamefont{Fantoni and Rosati}(1975)}]{FR75}
\bibinfo{author}{\bibfnamefont{S.}~\bibnamefont{Fantoni}} \bibnamefont{and}
  \bibinfo{author}{\bibfnamefont{S.}~\bibnamefont{Rosati}},
  \bibinfo{journal}{Nuovo Cimento A} \textbf{\bibinfo{volume}{25}},
  \bibinfo{pages}{593} (\bibinfo{year}{1975}).

\bibitem[{\citenamefont{Pandharipande and Wiringa}(1979)}]{PW79}
\bibinfo{author}{\bibfnamefont{V.~R.} \bibnamefont{Pandharipande}}
  \bibnamefont{and} \bibinfo{author}{\bibfnamefont{R.~B.}
  \bibnamefont{Wiringa}}, \bibinfo{journal}{Rev. Mod. Phys.}
  \textbf{\bibinfo{volume}{51}}, \bibinfo{pages}{821} (\bibinfo{year}{1979}),
  \urlprefix\url{https://link.aps.org/doi/10.1103/RevModPhys.51.821}.

\bibitem[{\citenamefont{Piarulli et~al.}(2019)\citenamefont{Piarulli, Bombaci,
  Logoteta, Lovato, and Wiringa}}]{Piarulli:2019pfq}
\bibinfo{author}{\bibfnamefont{M.}~\bibnamefont{Piarulli}},
  \bibinfo{author}{\bibfnamefont{I.}~\bibnamefont{Bombaci}},
  \bibinfo{author}{\bibfnamefont{D.}~\bibnamefont{Logoteta}},
  \bibinfo{author}{\bibfnamefont{A.}~\bibnamefont{Lovato}}, \bibnamefont{and}
  \bibinfo{author}{\bibfnamefont{R.~B.} \bibnamefont{Wiringa}}
  (\bibinfo{year}{2019}), \eprint{1908.04426}.

\bibitem[{\citenamefont{Bombaci and Logoteta}(2018)}]{Bombaci:2018ksa}
\bibinfo{author}{\bibfnamefont{I.}~\bibnamefont{Bombaci}} \bibnamefont{and}
  \bibinfo{author}{\bibfnamefont{D.}~\bibnamefont{Logoteta}},
  \bibinfo{journal}{Astron. Astrophys.} \textbf{\bibinfo{volume}{609}},
  \bibinfo{pages}{A128} (\bibinfo{year}{2018}), \eprint{1805.11846}.

\bibitem[{\citenamefont{Park et~al.}(1993)\citenamefont{Park, Min, and
  Rho}}]{Park:1993jf}
\bibinfo{author}{\bibfnamefont{T.-S.} \bibnamefont{Park}},
  \bibinfo{author}{\bibfnamefont{D.-P.} \bibnamefont{Min}}, \bibnamefont{and}
  \bibinfo{author}{\bibfnamefont{M.}~\bibnamefont{Rho}},
  \bibinfo{journal}{Phys. Rept.} \textbf{\bibinfo{volume}{233}},
  \bibinfo{pages}{341} (\bibinfo{year}{1993}), \eprint{hep-ph/9301295}.

\bibitem[{\citenamefont{Krebs et~al.}(2017)\citenamefont{Krebs, Epelbaum, and
  Meißner}}]{Krebs:2016rqz}
\bibinfo{author}{\bibfnamefont{H.}~\bibnamefont{Krebs}},
  \bibinfo{author}{\bibfnamefont{E.}~\bibnamefont{Epelbaum}}, \bibnamefont{and}
  \bibinfo{author}{\bibfnamefont{U.~G.} \bibnamefont{Meißner}},
  \bibinfo{journal}{Annals Phys.} \textbf{\bibinfo{volume}{378}},
  \bibinfo{pages}{317} (\bibinfo{year}{2017}), \eprint{1610.03569}.

\bibitem[{\citenamefont{Pastore et~al.}(2008)\citenamefont{Pastore, Schiavilla,
  and Goity}}]{Pastore:2008ui}
\bibinfo{author}{\bibfnamefont{S.}~\bibnamefont{Pastore}},
  \bibinfo{author}{\bibfnamefont{R.}~\bibnamefont{Schiavilla}},
  \bibnamefont{and} \bibinfo{author}{\bibfnamefont{J.~L.} \bibnamefont{Goity}},
  \bibinfo{journal}{Phys. Rev.} \textbf{\bibinfo{volume}{C78}},
  \bibinfo{pages}{064002} (\bibinfo{year}{2008}), \eprint{0810.1941}.

\bibitem[{\citenamefont{Pastore et~al.}(2009)\citenamefont{Pastore, Girlanda,
  Schiavilla, Viviani, and Wiringa}}]{Pastore:2009is}
\bibinfo{author}{\bibfnamefont{S.}~\bibnamefont{Pastore}},
  \bibinfo{author}{\bibfnamefont{L.}~\bibnamefont{Girlanda}},
  \bibinfo{author}{\bibfnamefont{R.}~\bibnamefont{Schiavilla}},
  \bibinfo{author}{\bibfnamefont{M.}~\bibnamefont{Viviani}}, \bibnamefont{and}
  \bibinfo{author}{\bibfnamefont{R.~B.} \bibnamefont{Wiringa}},
  \bibinfo{journal}{Phys. Rev.} \textbf{\bibinfo{volume}{C80}},
  \bibinfo{pages}{034004} (\bibinfo{year}{2009}), \eprint{0906.1800}.

\bibitem[{\citenamefont{Pastore et~al.}(2011)\citenamefont{Pastore, Girlanda,
  Schiavilla, and Viviani}}]{Pastore:2011ip}
\bibinfo{author}{\bibfnamefont{S.}~\bibnamefont{Pastore}},
  \bibinfo{author}{\bibfnamefont{L.}~\bibnamefont{Girlanda}},
  \bibinfo{author}{\bibfnamefont{R.}~\bibnamefont{Schiavilla}},
  \bibnamefont{and} \bibinfo{author}{\bibfnamefont{M.}~\bibnamefont{Viviani}},
  \bibinfo{journal}{Phys. Rev.} \textbf{\bibinfo{volume}{C84}},
  \bibinfo{pages}{024001} (\bibinfo{year}{2011}), \eprint{1106.4539}.

\bibitem[{\citenamefont{Kolling et~al.}(2009)\citenamefont{Kolling, Epelbaum,
  Krebs, and Meissner}}]{Kolling:2009iq}
\bibinfo{author}{\bibfnamefont{S.}~\bibnamefont{Kolling}},
  \bibinfo{author}{\bibfnamefont{E.}~\bibnamefont{Epelbaum}},
  \bibinfo{author}{\bibfnamefont{H.}~\bibnamefont{Krebs}}, \bibnamefont{and}
  \bibinfo{author}{\bibfnamefont{U.~G.} \bibnamefont{Meissner}},
  \bibinfo{journal}{Phys. Rev.} \textbf{\bibinfo{volume}{C80}},
  \bibinfo{pages}{045502} (\bibinfo{year}{2009}), \eprint{0907.3437}.

\bibitem[{\citenamefont{Kolling et~al.}(2011)\citenamefont{Kolling, Epelbaum,
  Krebs, and Meissner}}]{Kolling:2011mt}
\bibinfo{author}{\bibfnamefont{S.}~\bibnamefont{Kolling}},
  \bibinfo{author}{\bibfnamefont{E.}~\bibnamefont{Epelbaum}},
  \bibinfo{author}{\bibfnamefont{H.}~\bibnamefont{Krebs}}, \bibnamefont{and}
  \bibinfo{author}{\bibfnamefont{U.~G.} \bibnamefont{Meissner}},
  \bibinfo{journal}{Phys. Rev.} \textbf{\bibinfo{volume}{C84}},
  \bibinfo{pages}{054008} (\bibinfo{year}{2011}), \eprint{1107.0602}.

\bibitem[{\citenamefont{Krebs et~al.}(2020)\citenamefont{Krebs, Epelbaum, and
  Meißner}}]{Krebs:2020rms}
\bibinfo{author}{\bibfnamefont{H.}~\bibnamefont{Krebs}},
  \bibinfo{author}{\bibfnamefont{E.}~\bibnamefont{Epelbaum}}, \bibnamefont{and}
  \bibinfo{author}{\bibfnamefont{U.-G.} \bibnamefont{Meißner}}
  (\bibinfo{year}{2020}), \eprint{2001.03904}.

\bibitem[{\citenamefont{Bacca and Pastore}(2014)}]{Bacca:2014tla}
\bibinfo{author}{\bibfnamefont{S.}~\bibnamefont{Bacca}} \bibnamefont{and}
  \bibinfo{author}{\bibfnamefont{S.}~\bibnamefont{Pastore}},
  \bibinfo{journal}{J. Phys.} \textbf{\bibinfo{volume}{G41}},
  \bibinfo{pages}{123002} (\bibinfo{year}{2014}), \eprint{1407.3490}.

\bibitem[{\citenamefont{Piarulli et~al.}(2013)\citenamefont{Piarulli, Girlanda,
  Marcucci, Pastore, Schiavilla, and Viviani}}]{Piarulli:2012bn}
\bibinfo{author}{\bibfnamefont{M.}~\bibnamefont{Piarulli}},
  \bibinfo{author}{\bibfnamefont{L.}~\bibnamefont{Girlanda}},
  \bibinfo{author}{\bibfnamefont{L.~E.} \bibnamefont{Marcucci}},
  \bibinfo{author}{\bibfnamefont{S.}~\bibnamefont{Pastore}},
  \bibinfo{author}{\bibfnamefont{R.}~\bibnamefont{Schiavilla}},
  \bibnamefont{and} \bibinfo{author}{\bibfnamefont{M.}~\bibnamefont{Viviani}},
  \bibinfo{journal}{Phys. Rev.} \textbf{\bibinfo{volume}{C87}},
  \bibinfo{pages}{014006} (\bibinfo{year}{2013}), \eprint{1212.1105}.

\bibitem[{\citenamefont{Schiavilla et~al.}(2019)\citenamefont{Schiavilla,
  Baroni, Pastore, Piarulli, Girlanda, Kievsky, Lovato, Marcucci, Pieper,
  Viviani et~al.}}]{Schiavilla:2018udt}
\bibinfo{author}{\bibfnamefont{R.}~\bibnamefont{Schiavilla}},
  \bibinfo{author}{\bibfnamefont{A.}~\bibnamefont{Baroni}},
  \bibinfo{author}{\bibfnamefont{S.}~\bibnamefont{Pastore}},
  \bibinfo{author}{\bibfnamefont{M.}~\bibnamefont{Piarulli}},
  \bibinfo{author}{\bibfnamefont{L.}~\bibnamefont{Girlanda}},
  \bibinfo{author}{\bibfnamefont{A.}~\bibnamefont{Kievsky}},
  \bibinfo{author}{\bibfnamefont{A.}~\bibnamefont{Lovato}},
  \bibinfo{author}{\bibfnamefont{L.~E.} \bibnamefont{Marcucci}},
  \bibinfo{author}{\bibfnamefont{S.~C.} \bibnamefont{Pieper}},
  \bibinfo{author}{\bibfnamefont{M.}~\bibnamefont{Viviani}},
  \bibnamefont{et~al.}, \bibinfo{journal}{Phys. Rev. C}
  \textbf{\bibinfo{volume}{99}}, \bibinfo{pages}{034005}
  (\bibinfo{year}{2019}),
  \urlprefix\url{https://link.aps.org/doi/10.1103/PhysRevC.99.034005}.

\bibitem[{\citenamefont{Nevo~Dinur et~al.}(2019)\citenamefont{Nevo~Dinur,
  Hernandez, Bacca, Barnea, Ji, Pastore, Piarulli, and
  Wiringa}}]{NevoDinur:2018hdo}
\bibinfo{author}{\bibfnamefont{N.}~\bibnamefont{Nevo~Dinur}},
  \bibinfo{author}{\bibfnamefont{O.~J.} \bibnamefont{Hernandez}},
  \bibinfo{author}{\bibfnamefont{S.}~\bibnamefont{Bacca}},
  \bibinfo{author}{\bibfnamefont{N.}~\bibnamefont{Barnea}},
  \bibinfo{author}{\bibfnamefont{C.}~\bibnamefont{Ji}},
  \bibinfo{author}{\bibfnamefont{S.}~\bibnamefont{Pastore}},
  \bibinfo{author}{\bibfnamefont{M.}~\bibnamefont{Piarulli}}, \bibnamefont{and}
  \bibinfo{author}{\bibfnamefont{R.~B.} \bibnamefont{Wiringa}},
  \bibinfo{journal}{Phys. Rev.} \textbf{\bibinfo{volume}{C99}},
  \bibinfo{pages}{034004} (\bibinfo{year}{2019}), \eprint{1812.10261}.

\bibitem[{\citenamefont{Marcucci et~al.}(2016)\citenamefont{Marcucci, Gross,
  Pena, Piarulli, Schiavilla, Sick, Stadler, Van~Orden, and
  Viviani}}]{Marcucci:2015rca}
\bibinfo{author}{\bibfnamefont{L.~E.} \bibnamefont{Marcucci}},
  \bibinfo{author}{\bibfnamefont{F.}~\bibnamefont{Gross}},
  \bibinfo{author}{\bibfnamefont{M.~T.} \bibnamefont{Pena}},
  \bibinfo{author}{\bibfnamefont{M.}~\bibnamefont{Piarulli}},
  \bibinfo{author}{\bibfnamefont{R.}~\bibnamefont{Schiavilla}},
  \bibinfo{author}{\bibfnamefont{I.}~\bibnamefont{Sick}},
  \bibinfo{author}{\bibfnamefont{A.}~\bibnamefont{Stadler}},
  \bibinfo{author}{\bibfnamefont{J.~W.} \bibnamefont{Van~Orden}},
  \bibnamefont{and} \bibinfo{author}{\bibfnamefont{M.}~\bibnamefont{Viviani}},
  \bibinfo{journal}{J. Phys.} \textbf{\bibinfo{volume}{G43}},
  \bibinfo{pages}{023002} (\bibinfo{year}{2016}), \eprint{1504.05063}.

\bibitem[{\citenamefont{Girlanda et~al.}(2010)\citenamefont{Girlanda, Kievsky,
  Marcucci, Pastore, Schiavilla, and Viviani}}]{Girlanda:2010vm}
\bibinfo{author}{\bibfnamefont{L.}~\bibnamefont{Girlanda}},
  \bibinfo{author}{\bibfnamefont{A.}~\bibnamefont{Kievsky}},
  \bibinfo{author}{\bibfnamefont{L.~E.} \bibnamefont{Marcucci}},
  \bibinfo{author}{\bibfnamefont{S.}~\bibnamefont{Pastore}},
  \bibinfo{author}{\bibfnamefont{R.}~\bibnamefont{Schiavilla}},
  \bibnamefont{and} \bibinfo{author}{\bibfnamefont{M.}~\bibnamefont{Viviani}},
  \bibinfo{journal}{Phys. Rev. Lett.} \textbf{\bibinfo{volume}{105}},
  \bibinfo{pages}{232502} (\bibinfo{year}{2010}), \eprint{1008.0356}.

\bibitem[{\citenamefont{Pastore et~al.}(2013)\citenamefont{Pastore, Pieper,
  Schiavilla, and Wiringa}}]{Pastore:2012rp}
\bibinfo{author}{\bibfnamefont{S.}~\bibnamefont{Pastore}},
  \bibinfo{author}{\bibfnamefont{S.~C.} \bibnamefont{Pieper}},
  \bibinfo{author}{\bibfnamefont{R.}~\bibnamefont{Schiavilla}},
  \bibnamefont{and} \bibinfo{author}{\bibfnamefont{R.~B.}
  \bibnamefont{Wiringa}}, \bibinfo{journal}{Phys. Rev.}
  \textbf{\bibinfo{volume}{C87}}, \bibinfo{pages}{035503}
  (\bibinfo{year}{2013}), \eprint{1212.3375}.

\bibitem[{\citenamefont{Datar et~al.}(2013)}]{Datar:2013pbd}
\bibinfo{author}{\bibfnamefont{V.~M.} \bibnamefont{Datar}}
  \bibnamefont{et~al.}, \bibinfo{journal}{Phys. Rev. Lett.}
  \textbf{\bibinfo{volume}{111}}, \bibinfo{pages}{062502}
  (\bibinfo{year}{2013}), \eprint{1305.1094}.

\bibitem[{\citenamefont{Pastore et~al.}(2014)\citenamefont{Pastore, Wiringa,
  Pieper, and Schiavilla}}]{Pastore:2014oda}
\bibinfo{author}{\bibfnamefont{S.}~\bibnamefont{Pastore}},
  \bibinfo{author}{\bibfnamefont{R.~B.} \bibnamefont{Wiringa}},
  \bibinfo{author}{\bibfnamefont{S.~C.} \bibnamefont{Pieper}},
  \bibnamefont{and}
  \bibinfo{author}{\bibfnamefont{R.}~\bibnamefont{Schiavilla}},
  \bibinfo{journal}{Phys. Rev.} \textbf{\bibinfo{volume}{C90}},
  \bibinfo{pages}{024321} (\bibinfo{year}{2014}), \eprint{1406.2343}.

\bibitem[{\citenamefont{Pastore et~al.}(2019)\citenamefont{Pastore, Carlson,
  Gandolfi, Schiavilla, and Wiringa}}]{Pastore:2019urn}
\bibinfo{author}{\bibfnamefont{S.}~\bibnamefont{Pastore}},
  \bibinfo{author}{\bibfnamefont{J.}~\bibnamefont{Carlson}},
  \bibinfo{author}{\bibfnamefont{S.}~\bibnamefont{Gandolfi}},
  \bibinfo{author}{\bibfnamefont{R.}~\bibnamefont{Schiavilla}},
  \bibnamefont{and} \bibinfo{author}{\bibfnamefont{R.~B.}
  \bibnamefont{Wiringa}} (\bibinfo{year}{2019}), \eprint{1909.06400}.

\bibitem[{\citenamefont{Pastore
  et~al.}(2018{\natexlab{b}})\citenamefont{Pastore, Carlson, Cirigliano,
  Dekens, Mereghetti, and Wiringa}}]{Pastore:2017ofx}
\bibinfo{author}{\bibfnamefont{S.}~\bibnamefont{Pastore}},
  \bibinfo{author}{\bibfnamefont{J.}~\bibnamefont{Carlson}},
  \bibinfo{author}{\bibfnamefont{V.}~\bibnamefont{Cirigliano}},
  \bibinfo{author}{\bibfnamefont{W.}~\bibnamefont{Dekens}},
  \bibinfo{author}{\bibfnamefont{E.}~\bibnamefont{Mereghetti}},
  \bibnamefont{and} \bibinfo{author}{\bibfnamefont{R.~B.}
  \bibnamefont{Wiringa}}, \bibinfo{journal}{Phys. Rev.}
  \textbf{\bibinfo{volume}{C97}}, \bibinfo{pages}{014606}
  (\bibinfo{year}{2018}{\natexlab{b}}), \eprint{1710.05026}.

\bibitem[{\citenamefont{Cirigliano et~al.}(2018)\citenamefont{Cirigliano,
  Dekens, De~Vries, Graesser, Mereghetti, Pastore, and
  Van~Kolck}}]{Cirigliano:2018hja}
\bibinfo{author}{\bibfnamefont{V.}~\bibnamefont{Cirigliano}},
  \bibinfo{author}{\bibfnamefont{W.}~\bibnamefont{Dekens}},
  \bibinfo{author}{\bibfnamefont{J.}~\bibnamefont{De~Vries}},
  \bibinfo{author}{\bibfnamefont{M.~L.} \bibnamefont{Graesser}},
  \bibinfo{author}{\bibfnamefont{E.}~\bibnamefont{Mereghetti}},
  \bibinfo{author}{\bibfnamefont{S.}~\bibnamefont{Pastore}}, \bibnamefont{and}
  \bibinfo{author}{\bibfnamefont{U.}~\bibnamefont{Van~Kolck}},
  \bibinfo{journal}{Phys. Rev. Lett.} \textbf{\bibinfo{volume}{120}},
  \bibinfo{pages}{202001} (\bibinfo{year}{2018}), \eprint{1802.10097}.

\bibitem[{\citenamefont{Wang et~al.}(2019)\citenamefont{Wang, Hayes, Carlson,
  Dong, Mereghetti, Pastore, and Wiringa}}]{Wang:2019hjy}
\bibinfo{author}{\bibfnamefont{X.~B.} \bibnamefont{Wang}},
  \bibinfo{author}{\bibfnamefont{A.~C.} \bibnamefont{Hayes}},
  \bibinfo{author}{\bibfnamefont{J.}~\bibnamefont{Carlson}},
  \bibinfo{author}{\bibfnamefont{G.~X.} \bibnamefont{Dong}},
  \bibinfo{author}{\bibfnamefont{E.}~\bibnamefont{Mereghetti}},
  \bibinfo{author}{\bibfnamefont{S.}~\bibnamefont{Pastore}}, \bibnamefont{and}
  \bibinfo{author}{\bibfnamefont{R.~B.} \bibnamefont{Wiringa}},
  \bibinfo{journal}{Phys. Lett.} \textbf{\bibinfo{volume}{B798}},
  \bibinfo{pages}{134974} (\bibinfo{year}{2019}), \eprint{1906.06662}.

\bibitem[{\citenamefont{Tilley et~al.}(2002)\citenamefont{Tilley, Cheves,
  Godwin, Hale, Hofmann, Kelley, Sheu, and Weller}}]{Tilley:2002vg}
\bibinfo{author}{\bibfnamefont{D.~R.} \bibnamefont{Tilley}},
  \bibinfo{author}{\bibfnamefont{C.~M.} \bibnamefont{Cheves}},
  \bibinfo{author}{\bibfnamefont{J.~L.} \bibnamefont{Godwin}},
  \bibinfo{author}{\bibfnamefont{G.~M.} \bibnamefont{Hale}},
  \bibinfo{author}{\bibfnamefont{H.~M.} \bibnamefont{Hofmann}},
  \bibinfo{author}{\bibfnamefont{J.~H.} \bibnamefont{Kelley}},
  \bibinfo{author}{\bibfnamefont{C.~G.} \bibnamefont{Sheu}}, \bibnamefont{and}
  \bibinfo{author}{\bibfnamefont{H.~R.} \bibnamefont{Weller}},
  \bibinfo{journal}{Nucl. Phys.} \textbf{\bibinfo{volume}{A708}},
  \bibinfo{pages}{3} (\bibinfo{year}{2002}).

\bibitem[{\citenamefont{Tilley et~al.}(2004)\citenamefont{Tilley, Kelley,
  Godwin, Millener, Purcell, Sheu, and Weller}}]{Tilley:2004zz}
\bibinfo{author}{\bibfnamefont{D.~R.} \bibnamefont{Tilley}},
  \bibinfo{author}{\bibfnamefont{J.~H.} \bibnamefont{Kelley}},
  \bibinfo{author}{\bibfnamefont{J.~L.} \bibnamefont{Godwin}},
  \bibinfo{author}{\bibfnamefont{D.~J.} \bibnamefont{Millener}},
  \bibinfo{author}{\bibfnamefont{J.~E.} \bibnamefont{Purcell}},
  \bibinfo{author}{\bibfnamefont{C.~G.} \bibnamefont{Sheu}}, \bibnamefont{and}
  \bibinfo{author}{\bibfnamefont{H.~R.} \bibnamefont{Weller}},
  \bibinfo{journal}{Nucl. Phys.} \textbf{\bibinfo{volume}{A745}},
  \bibinfo{pages}{155} (\bibinfo{year}{2004}).

\bibitem[{\citenamefont{Knecht et~al.}(2012)}]{Knecht:2012fs}
\bibinfo{author}{\bibfnamefont{A.}~\bibnamefont{Knecht}} \bibnamefont{et~al.},
  \bibinfo{journal}{Phys. Rev.} \textbf{\bibinfo{volume}{C86}},
  \bibinfo{pages}{035506} (\bibinfo{year}{2012}).

\bibitem[{\citenamefont{Suzuki et~al.}(2003)\citenamefont{Suzuki, Fujimoto, and
  Otsuka}}]{Suzuki:2003}
\bibinfo{author}{\bibfnamefont{T.}~\bibnamefont{Suzuki}},
  \bibinfo{author}{\bibfnamefont{R.}~\bibnamefont{Fujimoto}}, \bibnamefont{and}
  \bibinfo{author}{\bibfnamefont{T.}~\bibnamefont{Otsuka}},
  \bibinfo{journal}{Phys. Rev. C} \textbf{\bibinfo{volume}{67}},
  \bibinfo{pages}{044302} (\bibinfo{year}{2003}),
  \urlprefix\url{https://link.aps.org/doi/10.1103/PhysRevC.67.044302}.

\bibitem[{\citenamefont{Chou et~al.}(1993)\citenamefont{Chou, Warburton, and
  Brown}}]{Chou:1993zz}
\bibinfo{author}{\bibfnamefont{W.~T.} \bibnamefont{Chou}},
  \bibinfo{author}{\bibfnamefont{E.~K.} \bibnamefont{Warburton}},
  \bibnamefont{and} \bibinfo{author}{\bibfnamefont{B.~A.} \bibnamefont{Brown}},
  \bibinfo{journal}{Phys. Rev.} \textbf{\bibinfo{volume}{C47}},
  \bibinfo{pages}{163} (\bibinfo{year}{1993}).

\bibitem[{\citenamefont{Warburton}(1986)}]{1986WA01}
\bibinfo{author}{\bibfnamefont{E.~K.} \bibnamefont{Warburton}},
  \bibinfo{journal}{Phys. Rev. C} \textbf{\bibinfo{volume}{33}},
  \bibinfo{pages}{303} (\bibinfo{year}{1986}),
  \urlprefix\url{https://link.aps.org/doi/10.1103/PhysRevC.33.303}.

\bibitem[{\citenamefont{Pritychenko et~al.}(2011)\citenamefont{Pritychenko,
  Běták, Kellett, Singh, and Totans}}]{1989BA31}
\bibinfo{author}{\bibfnamefont{B.}~\bibnamefont{Pritychenko}},
  \bibinfo{author}{\bibfnamefont{E.}~\bibnamefont{Běták}},
  \bibinfo{author}{\bibfnamefont{M.}~\bibnamefont{Kellett}},
  \bibinfo{author}{\bibfnamefont{B.}~\bibnamefont{Singh}}, \bibnamefont{and}
  \bibinfo{author}{\bibfnamefont{J.}~\bibnamefont{Totans}},
  \bibinfo{journal}{Nuclear Instruments and Methods in Physics Research Section
  A: Accelerators, Spectrometers, Detectors and Associated Equipment}
  \textbf{\bibinfo{volume}{640}}, \bibinfo{pages}{213–218}
  (\bibinfo{year}{2011}), ISSN \bibinfo{issn}{0168-9002},
  \urlprefix\url{http://www.sciencedirect.com/science/article/pii/S0168900211005584}.

\bibitem[{\citenamefont{Wiringa}(2006)}]{PhysRevC.73.034317}
\bibinfo{author}{\bibfnamefont{R.~B.} \bibnamefont{Wiringa}},
  \bibinfo{journal}{Phys. Rev. C} \textbf{\bibinfo{volume}{73}},
  \bibinfo{pages}{034317} (\bibinfo{year}{2006}),
  \urlprefix\url{https://link.aps.org/doi/10.1103/PhysRevC.73.034317}.

\bibitem[{\citenamefont{Wiringa
  et~al.}(2000{\natexlab{a}})\citenamefont{Wiringa, Pieper, Carlson, and
  Pandharipande}}]{WPCP00}
\bibinfo{author}{\bibfnamefont{R.~B.} \bibnamefont{Wiringa}},
  \bibinfo{author}{\bibfnamefont{S.~C.} \bibnamefont{Pieper}},
  \bibinfo{author}{\bibfnamefont{J.}~\bibnamefont{Carlson}}, \bibnamefont{and}
  \bibinfo{author}{\bibfnamefont{V.~R.} \bibnamefont{Pandharipande}},
  \bibinfo{journal}{Phys. Rev. C} \textbf{\bibinfo{volume}{62}},
  \bibinfo{pages}{014001} (\bibinfo{year}{2000}{\natexlab{a}}),
  \urlprefix\url{https://link.aps.org/doi/10.1103/PhysRevC.62.014001}.

\bibitem[{\citenamefont{Wiringa
  et~al.}(2000{\natexlab{b}})\citenamefont{Wiringa, Pieper, Carlson, and
  Pandharipande}}]{Wiringa:2000gb}
\bibinfo{author}{\bibfnamefont{R.~B.} \bibnamefont{Wiringa}},
  \bibinfo{author}{\bibfnamefont{S.~C.} \bibnamefont{Pieper}},
  \bibinfo{author}{\bibfnamefont{J.}~\bibnamefont{Carlson}}, \bibnamefont{and}
  \bibinfo{author}{\bibfnamefont{V.~R.} \bibnamefont{Pandharipande}},
  \bibinfo{journal}{Phys. Rev.} \textbf{\bibinfo{volume}{C62}},
  \bibinfo{pages}{014001} (\bibinfo{year}{2000}{\natexlab{b}}),
  \eprint{nucl-th/0002022}.

\bibitem[{\citenamefont{Hardy and Towner}(2015)}]{Hardy:2014qxa}
\bibinfo{author}{\bibfnamefont{J.~C.} \bibnamefont{Hardy}} \bibnamefont{and}
  \bibinfo{author}{\bibfnamefont{I.~S.} \bibnamefont{Towner}},
  \bibinfo{journal}{Phys. Rev.} \textbf{\bibinfo{volume}{C91}},
  \bibinfo{pages}{025501} (\bibinfo{year}{2015}), \eprint{1411.5987}.

\bibitem[{\citenamefont{Schiavilla et~al.}(1998)}]{Schiavilla:1998je}
\bibinfo{author}{\bibfnamefont{R.}~\bibnamefont{Schiavilla}}
  \bibnamefont{et~al.}, \bibinfo{journal}{Phys. Rev.}
  \textbf{\bibinfo{volume}{C58}}, \bibinfo{pages}{1263} (\bibinfo{year}{1998}),
  \eprint{nucl-th/9808010}.

\bibitem[{\citenamefont{Forest et~al.}(1996)\citenamefont{Forest,
  Pandharipande, Pieper, Wiringa, Schiavilla, and Arriaga}}]{Forest:1996kp}
\bibinfo{author}{\bibfnamefont{J.~L.} \bibnamefont{Forest}},
  \bibinfo{author}{\bibfnamefont{V.~R.} \bibnamefont{Pandharipande}},
  \bibinfo{author}{\bibfnamefont{S.~C.} \bibnamefont{Pieper}},
  \bibinfo{author}{\bibfnamefont{R.~B.} \bibnamefont{Wiringa}},
  \bibinfo{author}{\bibfnamefont{R.}~\bibnamefont{Schiavilla}},
  \bibnamefont{and} \bibinfo{author}{\bibfnamefont{A.}~\bibnamefont{Arriaga}},
  \bibinfo{journal}{Phys. Rev.} \textbf{\bibinfo{volume}{C54}},
  \bibinfo{pages}{646} (\bibinfo{year}{1996}), \eprint{nucl-th/9603035}.

\end{thebibliography}

\end{document}